\documentclass[journal]{IEEEtran}

\usepackage{listings}
\usepackage{caption}
\usepackage{xspace}
\usepackage{tabularx}
\usepackage{multirow}
\usepackage{booktabs}
\usepackage{graphicx}
\usepackage{colortbl}
\usepackage[ruled,linesnumbered]{algorithm2e}
\usepackage{epigraph}
\usepackage{listings}
\usepackage{tabularx}
\usepackage{rotating}
\usepackage{amssymb}
\usepackage{amsthm}
\usepackage{subfigure}
\usepackage{mathtools}
\usepackage{pifont}

\usepackage[export]{adjustbox}

\definecolor{Gray}{gray}{0.95}
\definecolor{eminence}{RGB}{108,48,130}
\definecolor{codegreen}{rgb}{0,0.6,0}
\definecolor{codegray}{rgb}{0.5,0.5,0.5}
\definecolor{codepurple}{rgb}{0.58,0,0.82}
\definecolor{backcolour}{rgb}{0.95,0.95,0.92}

\lstdefinestyle{classicstyle}{
  backgroundcolor=\color{backcolour}, commentstyle=\color{codegreen},
  keywordstyle=\color{magenta},
  numberstyle=\tiny\color{codegray},
  stringstyle=\color{codepurple},
  basicstyle=\ttfamily\footnotesize,
  breakatwhitespace=false,         
  breaklines=true,                 
  captionpos=b,                    
  keepspaces=true,                 
  numbers=left,                    
  numbersep=5pt,                  
  showspaces=false,                
  showstringspaces=false,
  showtabs=false,                  
  tabsize=2
}

\lstdefinestyle{interfaces}{
  float=tp,
  floatplacement=tbp,
  abovecaptionskip=-5pt
}

\newcommand{\lstbg}[3][0pt]{{\fboxsep#1\colorbox{#2}{\strut #3}}}
\lstdefinelanguage{diff}{
  basicstyle=\ttfamily\scriptsize,,
  morecomment=[f][\lstbg{red!20}]-,
  morecomment=[f][\lstbg{green!20}]+,
  morecomment=[f][\lstbg{yellow!20}]++,
  morecomment=[f][\textit]{@@},
  texcl=false
}

\lstdefinestyle{hs} {
    language = C,
    morekeywords = [3]{<<, >>},
    morekeywords = [4]{++},
    basicstyle=\ttfamily\scriptsize,
    commentstyle=\color{blue}\ttfamily,
    morecomment=[f][\lstbg{red!20}]-,
    morecomment=[f][\lstbg{green!20}]+,
    morecomment=[f][\lstbg{yellow!20}]++,
    morecomment=[f][\lstbg{yellow!20}]--,
    morecomment=[f][\textit]{@@},
    texcl=false
}

\newcommand{\tabincell}[2]{\begin{tabular}{@{}#1@{}}#2\end{tabular}}
\newcommand\cparagraph[1]{\vspace{1.2mm}\noindent \textbf{#1.}}

\begin{document}
%
\title{Resilient Charging Infrastructure via Decentralized Coordination of Electric Vehicles at Scale}

\author{\IEEEauthorblockN{Chuhao Qin\IEEEauthorrefmark{1}, Alexandru Sorici\IEEEauthorrefmark{2}, Andrei Olaru\IEEEauthorrefmark{2},
Evangelos Pournaras\IEEEauthorrefmark{1}, 
and Adina Magda Florea\IEEEauthorrefmark{2}
} \\

\IEEEauthorblockA{
\IEEEauthorrefmark{1}School of Computer Science, University of Leeds, UK\\
\IEEEauthorrefmark{2}AI-MAS Laboratory, National University of Science and Technology POLITEHNICA Bucharest, Romania
}

\thanks{}
}


%



\maketitle

\begin{abstract}
The rapid adoption of electric vehicles (EVs) introduces major challenges for decentralized charging control. Existing decentralized approaches efficiently coordinate a large number of EVs to select charging stations while reducing energy costs, preventing power peak and preserving driver privacy. However, they often struggle under severe contingencies, such as station outages or unexpected surges in charging requests. These situations create competition for limited charging slots, resulting in long queues and reduced driver comfort. To address these limitations, we propose a novel collective learning-based coordination framework that allows EVs to balance individual comfort on their selections against system-wide efficiency, i.e., the overall queues across all stations. In the framework, EVs are recommended for adaptive charging behaviors that shift priority between comfort and efficiency, achieving Pareto-optimal trade-offs under varying station capacities and dynamic spatio-temporal EV distribution. Experiments using real-world data from EVs and charging stations show that the proposed approach outperforms baseline methods, significantly reducing travel and queuing time. The results reveal that, under uncertain charging conditions, EV drivers that behave selfishly or altruistically at the right moments achieve shorter waiting time than those maintaining moderate behavior throughout. Our findings under high fractions of station outages and adversarial EVs further demonstrate improved resilience and trustworthiness of decentralized EV charging infrastructure.
\end{abstract}


%
\IEEEpeerreviewmaketitle

\section{Introduction}
\IEEEPARstart{E}{lectric} vehicles (EVs) are becoming a preferred option in intelligent transportation systems due to their energy efficiency and reduced emissions, critical in addressing environmental concerns and fuel shortages. According to recent global market reports, EV sales are projected to surpass 17 million units in 2024 (over $20\%$ market share), with over 20 million expected in 2025~\cite{IEA2025}. As governments expand public charging infrastructure to meet soaring demand, centralized charging management faces limitations in scalability, cost, and resilience (e.g., single points of failure)~\cite{zhang2016scalable,aravena2021decentralized}. A promising alternative lies in decentralized charging control among EVs. It aims to allow EVs to manage their charging based on local conditions, user preference and grid/station needs without a central authority. Proper decentralized charging control can promote a positive, flexible user experience and potentially optimize charging patterns. This leads to the reduction of energy costs and prevention of power peak or even blackout, supporting widespread EV adoption~\cite{patil2023integration}.

However, existing decentralized charging control approaches struggle to balance the individual driver comfort with overall system efficiency. The challenge becomes more pronounced under unexpected situations, such as station failures or sudden surges in charging demand, in which the number of EV requests exceeds the available charging slots. On the one hand, EVs coordinate to achieve collective goals, such as avoiding multiple vehicles competing for the same slot. Without such coordination, assigning hundreds or thousands of EVs to limited charging slots becomes inefficient, leading to longer queues~\cite{elghanam2024optimization}. On the other hand, EVs are often required to compromise their individual comfort to adhere to such coordination. Beyond queuing time, discomfort may include factors such as real-time location, battery state of charge, and range anxiety. These factors are inherently sensitive and often subject to privacy constraints, limiting the extent to which they can be shared or used for global optimization~\cite{liu2016optimal}. As a result, the quality of coordination is constrained by the limited visibility into individual vehicle states, making it difficult to achieve both high system efficiency and user satisfaction. 

\begin{figure}[!t]
    \centering
    \includegraphics[width=\linewidth]{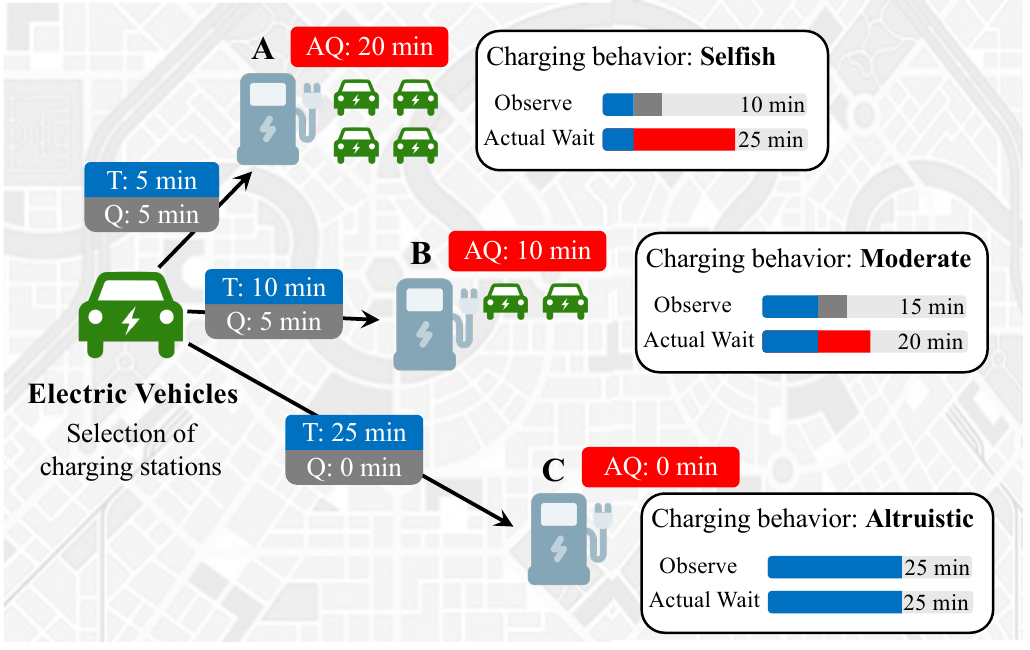}
    \caption{An example of \emph{DECharge} where an EV selects one of three charging stations. The EV observes the travel time ($T$) and estimated queuing time ($Q$) at each station, and chooses its behavior (selfish, moderate or altruistic) according to $T+Q$. After all EVs make their choices, the actual queuing time ($AQ$) at each station is determined via coordination. This coordination guides EVs toward moderate behavior, effectively reducing their actual waiting time.}
    \label{fig:preliminary}
\end{figure}

To bridge this research gap, we propose a new decentralized EV charging control framework, named \emph{DECharge}, \textit{\underline{D}iscomfort and  \underline{E}fficient-aware electric vehicle \underline{Charge} coordination} (Contribution 1). In this framework, as shown in Fig.~\ref{fig:preliminary}, EVs autonomously select charging stations from a set of available options, each representing a potential slot at a station. The core objective is to coordinate a large number of EVs to make choices by balancing \emph{driver discomfort} and \emph{system inefficiency}. Driver discomfort represents the individual cost of an option, estimated from factors including travel distance/time and observed queuing time, while system inefficiency indicates the overall queuing load and imbalances across all stations. \emph{DECharge} introduces the ``charging behavior'', i.e., allowing EVs to behave between selfish and altruistic during coordination. Selfish behaviors prioritize options with minimal driver discomfort, whereas altruistic behaviors accept a higher discomfort to reduce system inefficiency and alleviate the congestion at popular stations. The moderate behaviors balance both objectives instead. The coordination problem is modeled as a multi-objective, discrete-choice combinatorial optimization problem, which is NP-hard. Solving it in a decentralized manner demands a cost-effective approach capable of handling large-scale and efficient information exchange while adapting to complex and dynamic charging conditions, e.g., varying station capacities and spatio-temporal EV demand patterns~\cite{lehnhoff2019towards}.

Therefore, \emph{DECharge} employs collective learning~\cite{pournaras2018decentralized}, a state-of-the-art decentralized learning method for optimized selection on charging stations (Contribution 2). It is selected for its scalability (supporting large number of agents), efficiency (low computational and communication cost) and resilience~\cite{pournaras2018decentralized}. To further improve adaptability under real-time conditions, we introduce a charging behavior recommendation mechanism managed by a system or traffic operator (Contribution 3). This mechanism assists EVs on adopting contextually appropriate charging behaviors within defined time windows. By dividing the overall operation into discrete time windows, EVs are able to observe queuing patterns and spatial distribution of nearby vehicles in a timely manner. Based on these observations, each EV can dynamically adjust its strategy by choosing between selfish and altruistic behavior to better navigate the Pareto-optimal trade-off between driver discomfort and system inefficiency. For example, EVs appearing in areas with high vehicle density are recommended to choose distant stations (altruistic behavior). The open-source implementation of \emph{DECharge} serves as a benchmark for future research\footnote{Available at: https://github.com/TDI-Lab/DECharge.} (Contribution 4).

Finally, quantitative findings using real-world data demonstrates that \emph{DECharge} outperforms baseline methods in reducing both travel distance and queuing time of EVs. The results suggest that EVs adapt their charging strategies dynamically in response to varying station capacities and the spatio-temporal distribution of charging demand, leading to lower average waiting time compared to scenarios where all EVs adopt moderate behaviors. Experimental results further prove the resilience and trustworthiness of EV charging infrastructure: \emph{DECharge} balance the distribution of energy demand across stations, preventing the risk of power peaks. Furthermore, \emph{DECharge} maintains high performance even in the presence of a high ratio of station failures or a high fraction of EVs exhibiting adversary behavior (Contribution 5).

The rest of this paper is organized as follows. Section~\ref{sec:related} presents recent literature related to this study and explains how this paper goes beyond their approaches. Section~\ref{sec:model} describes the system model of \emph{DECharge} and the addressed problem, while the detailed system design of \emph{DECharge} is given in Section~\ref{sec:method}. Section~\ref{sec:settings} presents experimental settings using real-world data, and Section~\ref{sec:evaluation} evaluates experimental results of \emph{DECharge}. Finally, Section \ref{sec:conclusion} summarizes and concludes the paper.

\begin{table}
	\centering
	\caption{Comparison to related work: (\checkmark) indicates criteria covered, ($\times$) indicates criteria not covered.}  
	\label{table:criteria}
	\resizebox{\linewidth}{!}{
	\begin{tabular}{lcccccc}  
		\toprule  
		\textbf{Criteria \, Ref.:} &\tabincell{l}{\cite{zhang2020cddpg}} 
        &\tabincell{l}{\cite{zhang2021intelligent}} 
        &\tabincell{l}{\cite{da2019coordination}} 
        &\tabincell{l}{\cite{acs2024distributed}} 
        &\tabincell{l} {\cite{pournaras2019socio}} 
        &This work\\  
		\midrule  
		Decentralization  &$\times$ &\checkmark &\checkmark &\checkmark &\checkmark &\checkmark \\   
		Scalability		  &$\times$	&\checkmark	&\checkmark	&$\times$	&\checkmark &\checkmark \\
		Adaptability 	  &\checkmark &\checkmark &\checkmark	&$\times$	&$\times$	&\checkmark \\  
		Resilience		&$\times$	&$\times$  	&$\times$ 	&$\times$   &$\times$ 	&\checkmark	 \\
  	Privacy-preservation	&$\times$	&$\times$	&$\times$	&$\times$ &\checkmark	&\checkmark \\
		\bottomrule
	\end{tabular}  
	}
\end{table}

\section{Related Work} \label{sec:related}
Real-time charging control is critical for the efficient operation of EVs and has become a popular topic in intelligent transportation systems. Previous studies have applied heuristic approaches such as particle swarm optimization~\cite{ko2013optimal}, greedy algorithms~\cite{efthymiou2017electric}, ant colony optimization~\cite{mavrovouniotis2018ant}, and deep reinforcement learning~\cite{zhang2020cddpg} to derive charging control strategies. However, these methods rely on centralized computation, which can be costly, difficult to scale, and vulnerable to single points of failure.

More recently, research has shifted toward decentralized charging control~\cite{aravena2021decentralized,pustivsek2016blockchain,nimalsiri2019survey,nimalsiri2021coordinated,ping2020two}. For instance, Zhang \textit{et al.}~\cite{zhang2021intelligent} employed a multi-agent reinforcement learning framework with centralized training and decentralized execution to recommend publicly accessible charging stations while addressing spatio-temporal imbalances in EV demand. Azzouz \textit{et al.}~\cite{azzouz2023optimization} proposed a decentralized scheduling method in which each EV independently selects its optimal charging time slot according to driver preferences without a central coordinator. Zhu \textit{et al.}~\cite{zhu2024multi} developed a multi-objective combinatorial optimization problem for planning placement and capacity of charging stations to optimize the benefits for both operators and users by minimizing the construction, operation and maintenance, costs, and the user’s detour time. While these methods improve scalability and adaptability and effectively minimize operational costs such as travel distance, queuing time, and charging fees, they primarily focus on behaviors of charging stations. In contrast, the drivers' perspective or charging behaviors, i.e., capturing EV users’ range anxiety, waiting time, and comfort, requires to explore~\cite{elghanam2024optimization}.


A deep understanding of EV drivers’ charging behaviors is crucial for improving EV adoption, user satisfaction, and system trustworthiness~\cite{patil2023integration,yang2018investigation,helmus2020data,song2023learning,chen2024flexibility}. Earlier work has analyzed and modeled charging behaviors using multi-source data to estimate EV preferences for station selection~\cite{li2018planning,wang2021electric}. Silva \textit{et al.}~\cite{da2019coordination} applied multi-agent reinforcement learning to determine when an EV driver should act selfishly by immediately recharging its battery or act cooperatively by deferring charging to a more suitable time. In this setting, EVs rely only on local observations and limited communication, without requiring manual inputs regarding daily energy needs. However, these approaches still assume that drivers are willing to share sensitive private information such as battery levels and locations. 

In the review of decentralized heuristics, collective learning has emerged as a powerful approach for solving multi-objective, discrete-choice combinatorial optimization problems in a decentralized manner~\cite{pournaras2020holarchic,yahya2017collective}. It is a distributed learning approach where autonomous agents coordinate their decisions to collectively optimize tasks that benefit from cooperation, such as mitigating electric power peaks. Pournaras \textit{et al.}~\cite{pournaras2018decentralized} proposed I-EPOS, \textit{Iterative Economic Planning and Optimized Selections}, a decentralized coordination framework where agents exchange information over a tree-structured communication topology. Compared to other well-known heuristics such as DSA, MGM, and COHDA~\cite{maheswaran2004distributed,acs2024distributed,hinrichs2014cohda}, agents in I-EPOS require no full information of the system, thereby significantly reducing both computational and communication overhead.  

A closely related line of work leverages I-EPOS for privacy-preserving decentralized EV charging control, addressing socio-technical aspects such as reliability, discomfort, and fairness in the Smart Grid~\cite{pournaras2019socio}. This fully participatory learning mechanism prevents sudden peaks in power demand while avoiding the need to acquire sensitive driver information. However, two key challenges remain: (i) This work overlook the problem of fully occupied charging slots and long queuing times caused by surges in charging requests; and (ii) it assumes that EV charging behaviors are the same, without allowing them to dynamically adapt their station selections according to real-time spatial distribution of EVs and station capacities.

Table~\ref{table:criteria} illustrates the comparison to related work across multiple criteria. Here, \textit{decentralization} indicates that the method coordinates EVs in a decentralized way; \textit{scalability} refers to the ability of the method to scale efficiently with low computational and communication overheads; \textit{adaptability} indicates whether EVs can adapt their decision-making on charging selection in respond to dynamic environments (e.g., changing spatio-temporal distribution of EVs); \textit{resilience} refers to the ability of the method to prevent single point of failure; \textit{privacy-preservation} indicates whether the method respects private information of EV drivers.

\begin{table}[!t]
	\caption{Notations.}  
	\centering
	\begin{tabularx}{\linewidth}{lXl}  
		\hline  
		Notation & Exoptionation \\  
		\hline    
		$ t, T $  & The index of a time window; total number of time windows \\  
		$ n, N  $  & The index of an EV charging request; total number of requests \\ 
        $ N_t$ & The set of requests at time window $t$ \\
        $ m, M $  & The index of a charging station; total number of stations \\ 
        $ x_{nm} $  & The binary decision variable which takes 1 if $n$ selects station $m$ and 0 otherwise \\
        $\beta_n$ & The charging behavior parameter of request $n$ \\
        $ d_n, l $  & The charging demand of request $n$; the current location \\
        $ t_n $ & The time that request $n$ starts \\
        $ \tau_{mj} $  & The slot free time of charging slot $j$ at station $m$ \\
        $\tau^{\text{Q}}_m $ & The actual queuing time of station $m$\\
        $ \mathbb{D}_{nm} $  & The driver discomfort if request $n$ selects station $m$ \\
		$ D_{nm} $  & Relative travel distance between request $n$ and station $m$ \\
        $ D^{\text{max}}_{n} $  & The maximum travel distance of request $n$\\
        $Q_{nm}$ & The queuing time of station $m$ observed by request $n$ \\
        $ \alpha_1, \alpha_2 $  & The constant parameters to normalize $D_{nm}$ and $Q_{nm}$ \\
        $ \mathcal{O}_1, \mathcal{O}_2 $  & The objective functions of driver discomfort and system inefficiency \\
        $k_{nm} $ & The charging option of request $n$ on station $m$  \\
        $\mathcal{K}_n$ & The set of charging options \\
        $G, B $ & The aggregated options; the branch response \\
        $I$ & The index of iteration in collective learning \\
        $f_g $ & The overall cost function \\
        $\delta_c $ & The agent's decision regarding whether to approve the charging selection of its child $c$ \\
        $J^\text{A}_n$ & Number of available slots within the range of request $n$ \\
        $\gamma, \omega_t$ & Constant factor to calculate charging behavior; The prediction coefficient to estimate the number of requests at $t+1$ \\
        \hline 
	\end{tabularx}  
	\label{table:notation}
\end{table}

\section{System Model} \label{sec:model}
This section introduces the proposed decentralized EV charging control framework \emph{DECharge}, which is modeled as a multi-objective, discrete-choice combinatorial optimization problem. We begin by defining the key components of the model, followed by a formal problem formulation based on two primary optimization objectives. Table~\ref{table:notation} lists the notations used in this paper.

\subsection{Definitions and assumptions}
Consider a number of $N$ charging requests of EVs and $M$ charging stations over a grid representing a 2D map over a long time span (e.g., an entire day). The time span is divided into a number of $T$ time windows with equal length, each time window represented as $t$, $t \leq T$. For example, a time window can be set from 10:00 to 10:30. The set of charging requests of EVs from $t-1$ to $t$ is defined as $N_t$, where $\sum_{t=1}^T |N_t| = N$. Additionally, we define the EV charging request and charging station as below.

\cparagraph{EV charging request}
A charging request of an EV, equipped with a software agent, can autonomously generate a number of possible charging options that determine which charging station it selects to charge. Each request is represented by $\{n, t_n, d_n, l_n\}$, where $n$ is the index of this charging request appeared in the time window $t$, $n \in N_t$; $t_n$ indicates the time that $n$ starts to request the charge; $d_n$ denotes the charging time demanded by $n$, which is determined by the battery state of charge; and $l_n$ is the current location of $n$. For each index of charging station $m$, $m \leq M$, the charging option of $n$ is denoted as $x_{nm}$, which is a binary variable which takes 1 if $n$ selects the charging station $m$ to charge and 0 otherwise, $m \leq M$. Note that each EV can only select one station. Here, the current location $l_n$ is private to $n$ such that they can hide their sensitive information, whereas the charging demand $d_n$ and selection $x_{nm}$ can be shared to other EVs.

\cparagraph{Charging station}
A charging station $m$ is defined as $\{m, l_m, \tau_{mj}, \tau^{\text{Q}}_m (t) \}$, where $m$ is the index of charging station; $l_m$ is the current location of $m$, $\tau_{mj}$ indicates the slot free time when the charging slot $j$ in $m$ becomes available, $j \leq J$; and $\tau^{\text{Q}}_m (t)$ denotes the actual queuing time, which is updated in every time window $t$. At the beginning $t = 0$, both slot free time and actual queuing time are set as $0$, $\tau_{mj} = \tau^{\text{Q}}_m (0) = 0$. Next, they are updated through Algorithm~\ref{algorithm1} within each time window. For all EV charging requests $N_t$ in time window $t$, the EV that requests earlier is assigned first, i.e., first-come, first-served queuing (Line 2). When a new EV $n$ selects station $m$, it is assigned to the charging slot that can complete the charging process in the shortest time (Line 4). The actual queuing time of the station is then increased only if the selected slot remains occupied when the new EV arrives (Line 5). Finally, the slot free time is also extended by the demand time $d_n$ (Line 6).

To simplify the analysis and focus on the core problem of decentralized EV charging coordination, several assumptions are made in the model design: (1) Traffic conditions and variations in EV travel speeds are not considered, and the travel speed is treated as constant. (2) All charging stations are assumed to have the same maximum number of available slots, allowing the study to emphasize coordination rather than infrastructure heterogeneity. (3) The charging demand time $d_n$ of each EV is assumed to remain constant and is unaffected by additional energy consumption incurred during travel or queuing, even though these factors reduce battery state of charge. (4) Potential communication issues among EVs, including delays, packet loss, or transmission errors, are not considered, assuming reliable message exchange. (5) Energy trading or charge-discharge interactions between EVs and charging stations are not modeled, as the focus is on on charging behavior optimization rather than energy market dynamics. Moreover, we assume each EV only selects a station within its predefined time window, and thus variations in power prices across peak/off-peak hours are not considered.

\begin{algorithm}[!t]
\SetAlgoLined
	\caption{State update of charging stations.} 
	\label{algorithm1}
	\KwIn{State of charging stations $\{ \tau_{mj} \}$, set of EV charging requests $N_t$, their demands $d_n$, and their charging options $x_{nm}$.}
    \KwOut{Updated state of the charging station $\{\tau_{mj}, \tau^{\text{Q}}_m (t) \}$.}
    $\tau^{\text{Q}}_m (t) = 0$\;
    Sort the $N_t$ by $t_n$\;
    \For{$n \in N_t$}
    {
        $j^* = \underset{j}{\text{argmin}} \; \tau_{mj} \cdot x_{nm}$\;
        $\tau^{\text{Q}}_m (t) = \tau^{\text{Q}}_m (t) + \max (\tau_{mj^*} - t_n, 0)$\;
        $\tau_{mj^*} = \max (\tau_{mj^*}, t_n) + d_n$\;
    }
\end{algorithm} 

\begin{figure}[!t]
    \centering
    \includegraphics[width=0.9\linewidth]{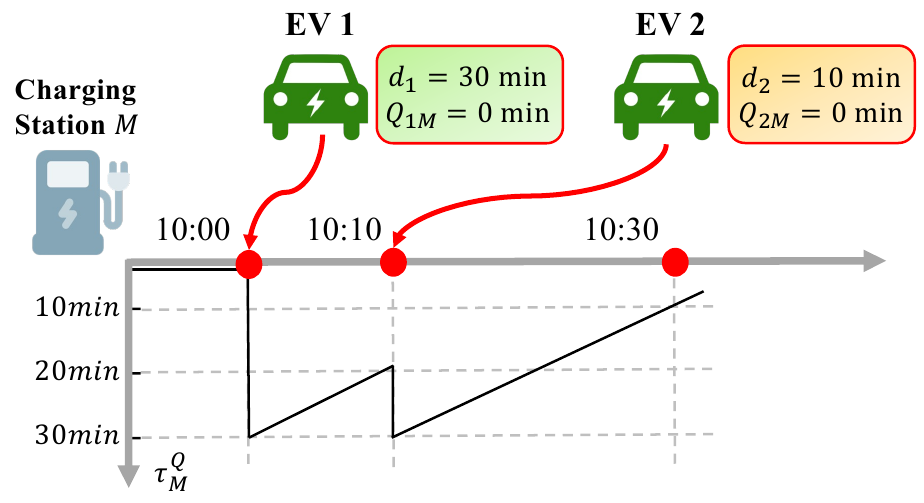}
    \caption{A case where two EVs choose the same charging station without coordination. EV 1 requests charging time of 30 min and EV 2 requests 10 min. They start to request charging at 10:00, but reach the station at 10:00 and 10:10 respectively. The coordinate axis shows the update of actual queuing time at charging station $M$.}
    \label{fig:scenario}
\end{figure}

\subsection{Problem formulation}
The core problem in decentralized EV charging control lies in \textit{coordinating multiple EVs within each time window to autonomously and collaboratively select optimal charging stations, aiming to minimize their operational cost}. This operational cost includes both travel distance to the station and queuing time upon arrival. During coordination, each EV must estimate the operational cost of its charging options and share its selected option with others at the beginning of the period. However, discrepancies can arise between observed and actual queuing time, as these estimates are made in parallel and independently, and do not account for conflicts in real-time decisions. For example, see Fig.~\ref{fig:scenario}, both EV 1 and EV 2 that have the same distance to station $M$ start to request charging at 10:00, such that they observe $0$ min queuing time at $M$. If both of them select this station without coordination, EV 2 arrives at the station later and thus it must wait $20$ min until EV 1 finish charging (first-come, first-served queuing). 

To tackle this problem, two key objectives are introduced: minimizing \textit{driver discomfort} and \textit{system inefficiency}. The first aims to allow each EV to privately estimate its operational cost without revealing sensitive information, while the second encourages collective behavior that minimizes actual queuing delays across the system. By jointly optimizing these objectives, the system can significantly reduce real waiting times in a decentralized and privacy-preserving fashion. The detailed formulations of these objectives are illustrated as follows:

\cparagraph{Driver discomfort} It denotes the estimated cost of an EV charging request when selecting a possible charging station, formulated as:
\begin{equation}
    \mathbb{D}_{nm} = \alpha_1 \cdot D_{nm} + \alpha_2 \cdot Q_{nm},
\end{equation}
where $D_{nm}$ indicates the relative travel distance from charging request $n$ to station $m$, $Q_{nm}$ is the queuing time of station $m$ observed by EV $n$, $\alpha_1$ and $\alpha_2$ are constant parameters to normalize travel distance and observed queuing time $\alpha_1, \alpha_2 \in [0, 1]$. These parameters are private to request $n$ and can be adjusted according to individual preference on traveling or queuing. The driver discomfort aims to minimize the expected waiting time of all charging requests, which is calculated by the weighted summation of both travel distance and observed queuing time. The first objective is to minimize the driver discomfort cost of each EV request, which is formulated as below:
\begin{equation}
    \mathcal{O}_1 (x_{nm}) = \sum_{m=1}^M \mathbb{D}_{nm} \cdot x_{nm}.
    \label{eq:obj_discomfort}
\end{equation}

\cparagraph{System inefficiency} It denotes the imbalance of total queuing time across station. Error and correlation metrics such as the root mean square error or variance can estimate this imbalance, which is shown earlier to be an NP-hard combinatorial optimization problem in this context~\cite{pournaras2018decentralized}. The system inefficiency aims to prevent excessive EVs requesting the same charging station, and ultimately decrease the actual queuing time over all time windows. Note that the actual queuing time is obtained only after all EVs are assigned to charging slots. The objective is to minimize the system inefficiency cost of charging stations at each time window, formulated as follows:
\begin{equation}
    \mathcal{O}_2 (x_{nm}) = \sqrt{\frac{\sum_{m=1}^M [\tau^{\text{Q}}_m (t)]^2}{M}}.
    \label{eq:obj_inefficiency}
\end{equation}

Based on the aforementioned formulations, this problem can be modeled as a NP-hard multi-objective discrete-choice combinatorial optimization problem, which is listed as follows:
\begin{equation}
    \min \; \{ \sum^N_{n=1} \mathcal{O}_1 (x_{nm}), \sum^T_{t=1} \mathcal{O}_2 (x_{nm})\},
    \label{eq:objs}
\end{equation}

Subject to 
\begin{equation}
    \sum_{m=1}^M x_{nm} = 1, \; \forall n \leq N,
    \label{eq:const_select}
\end{equation}
\begin{equation}
    D_{nm} \cdot x_{nm} \leq D^{\text{max}}_n, \forall n \leq N, \forall m \leq M,
    \label{eq:const_travel}
\end{equation}
\begin{equation}
    Q_{nm} = \tau^{\text{Q}}_m (t - 1), \; \forall n \in N_t, \forall m \leq M.
    \label{eq:const_queue}
\end{equation}

Constraint (\ref{eq:const_select}) restricts that each EV only chooses one charging station; Constraint (\ref{eq:const_travel}) limits the maximum travel distance of EVs according to their range anxiety that drivers concern about running out of battery; Constraint (\ref{eq:const_queue}) restricts that each EV that requests charging at a time window only observes the queuing time at the beginning of this window. 

To jointly minimize these two objectives Eq.(\ref{eq:obj_discomfort}), (\ref{eq:obj_inefficiency}), and achieve Pareto optimality, a weight parameter called \emph{charging behavior} is introduced for a weighted sum of objectives. The detailed definition of this parameter is illustrated as follows:

\cparagraph{Charging behavior} It indicates the EV driver preference that chooses a more selfish, altruistic or moderate behavior, denoted as $\beta_n$, $0 \leq \beta \leq 1$. As the value of $\beta_n$ increases, the EV becomes more selfish, prioritizing the choice of charging station with low driver discomfort at the expense of higher system inefficiency. Thus, the overall cost function can be formulated as follows:
\begin{equation}
    \mathop{\min}\limits_{x_{nm}}\; \sum^T_{t=1} \sum_{n \in N_t} [\beta_n \cdot \mathcal{O}_1 (x_{nm}) + (1 - \beta_n) \cdot \mathcal{O}_2 (x_{nm})],
    \label{eq:obj_overall}
\end{equation}
By adjusting the value of $\beta_n$, each EV can choose a charging option by prioritizing one of two contradictory objectives, i.e., achieving lower driver discomfort or lower system inefficiency. This contradiction is validated through theoretical analysis in Appendix~\ref{sec:theoretical}.

\begin{figure}[!t]
    \centering
    \includegraphics[width=\linewidth]{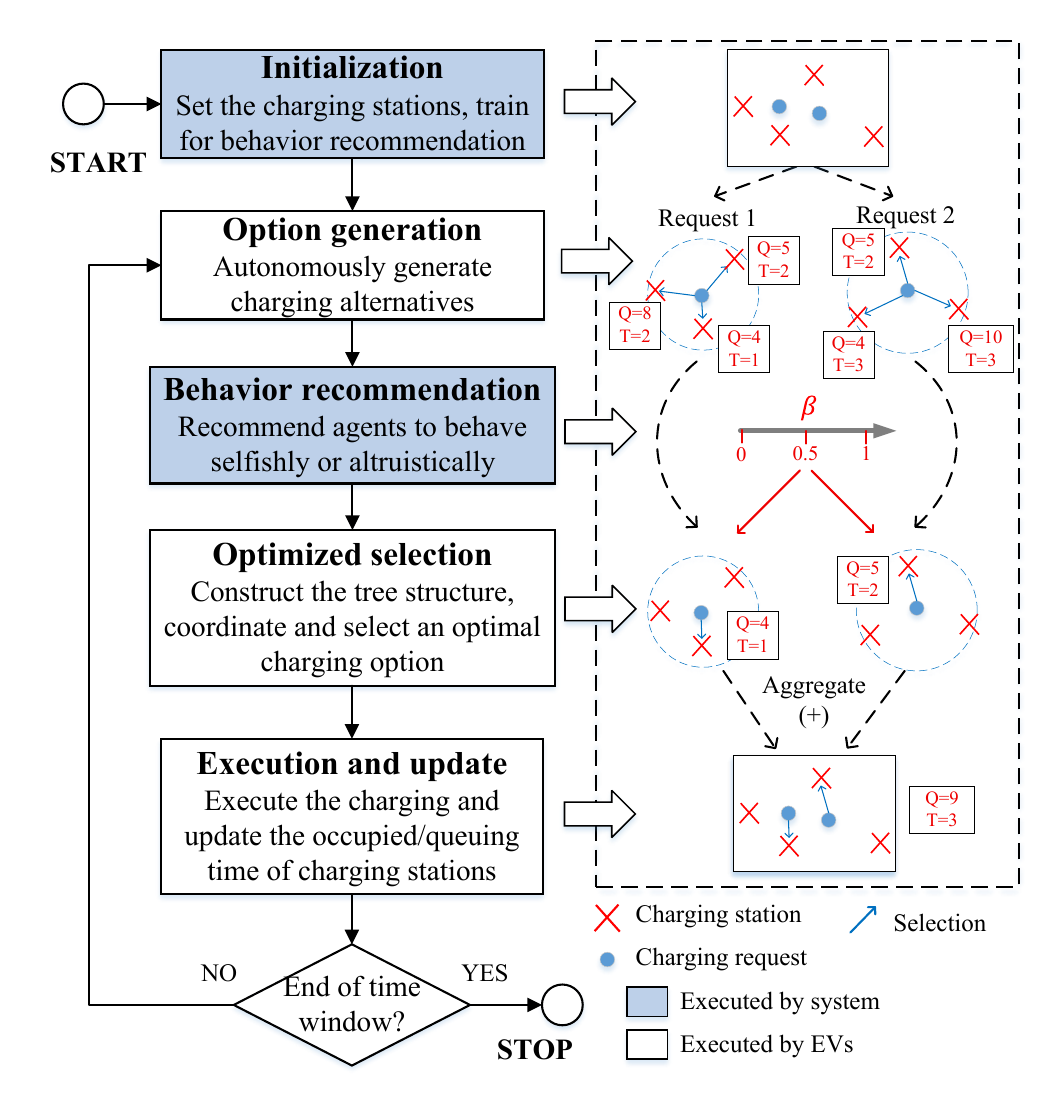}
    \caption{The overall system framework of \emph{DECharge} for decentralized EV charging control ($Q$ is the queuing time (minute) observed by EVs; $T$ is the travel distance (km) from EVs to stations).}
    \label{fig:framework}
\end{figure}

\section{Detailed System Design} \label{sec:method}
For the objectives of minimizing both driver discomfort and system inefficiency, \emph{DECharge} aims to find an approximate solution to the NP-hard problem by efficiently coordinating EVs to select an optimal charging option in a decentralized manner.

Fig.~\ref{fig:framework} shows the designed system framework of \emph{DECharge}, consisting of the following steps: Firstly, the environmental map is initialized by setting a number of charging stations with initialized occupied time on their slots. Meanwhile, the coefficients used in behavior recommendation are trained, which is illustrated in Section~\ref{sec:behavior_recommend}. At each time window, new EVs show up in the map (i.e., new charging requests) and autonomously generate their charging options, each indicating a possible choice of a charging station nearby. For each option, EVs observe their relative distance to the corresponding charging station and the queuing time. These pieces of information are used to calculate the driver discomfort. Next, these EVs are recommended to behave between selfish and altruistic (i.e., determining the value of charging behavior $\beta_n$). Then, each EV in the map interacts with others to autonomously select an optimized option from the generated alternatives through the coordination of a multi-agent collective learning (I-EPOS). This step aims to find the optimal combination of options selected by charging requests to minimize both driver discomfort and system inefficiency. After that, these EVs execute their selected charging options by charging at the stations whereas the occupied time of stations are updated through Algorithm~\ref{algorithm1}. In the next time window, new EVs emerge and continue executing the aforementioned steps. Finally, the system terminates if the time window ends.

The details of charging option generation and optimized selection are illustrated at the following sub-sections.

\begin{algorithm}[!t]
\SetAlgoLined
	\caption{Charging option generation of a EV charging request.} 
	\label{algorithm2}
	\KwIn{The state of charging request $\{ n, d_n, l_n \}$ and stations $\{ m, l_n, \tau^{\text{Q}}_m(t-1) \}$, $\forall m \leq M$.}
    \KwOut{The set of options $\mathcal{K}_n$ of $n$.}
	Initialize the set of options $\mathcal{K}_n = \emptyset$\;
    \For{charging station $m := 1,...,M$}
    {
        $D_{nm} = |l_n - l_m|$ \;
        \If{$D_{nm} \leq D^{\text{max}}_n$}
        {
            $Q_{nm} = \tau^{\text{Q}}_{m} (t-1)$ \;
            $\mathbb{D}_{nm} = \alpha_1 \cdot D_{nm} + \alpha_2 \cdot Q_{nm}$\;
            $k_{nm} = (x_{nm}, t_n, d_n, \mathbb{D}_{nm})$\;
            $\mathcal{K}_n \rightarrow k_{nm} \cup \mathcal{K}_n$\;
        }
    }
\end{algorithm} 

\subsection{Charging option generation strategy}
Algorithm~\ref{algorithm2} outlines the following steps of the charging option generation for each EV charging request $n$, $\forall n \leq N$: With the input of the information of EV charging request $\{ n, d_n, l_n \}$ and all stations located at the map $\{ m, l_n, \tau_{mj} (t), \tau^{\text{Q}}_m(t) \}$, $\forall m \leq M$, each EV aims to generate charging alternatives for stations and store them into a set $\mathcal{K}_n$ for an optimized selection. Firstly, for each of station, the EV computes the relative travel distance $D_{nm}$ from its current location $l_n$ to the station $l_m$ (Line 3). If $D_{nm}$ is no more than maximum travel distance, then $n$ observes the queuing time of $m$, which is $Q_{nm} = \tau^{\text{Q}}_{m} (t)$ (Line 5). Next, the request $n$ calculates the driver discomfort of this charging alternative, and generates an option with the index of $k$ based on its station selection $x_{nm}$, charging demand $d_n$, and driver discomfort $\mathbb{D}_{nm}$ (Lines 6-7). Finally, the generated option is added to $\mathcal{K}_n$ (Line 8).

\begin{figure}[!t]
    \centering
    \includegraphics[width=\linewidth]{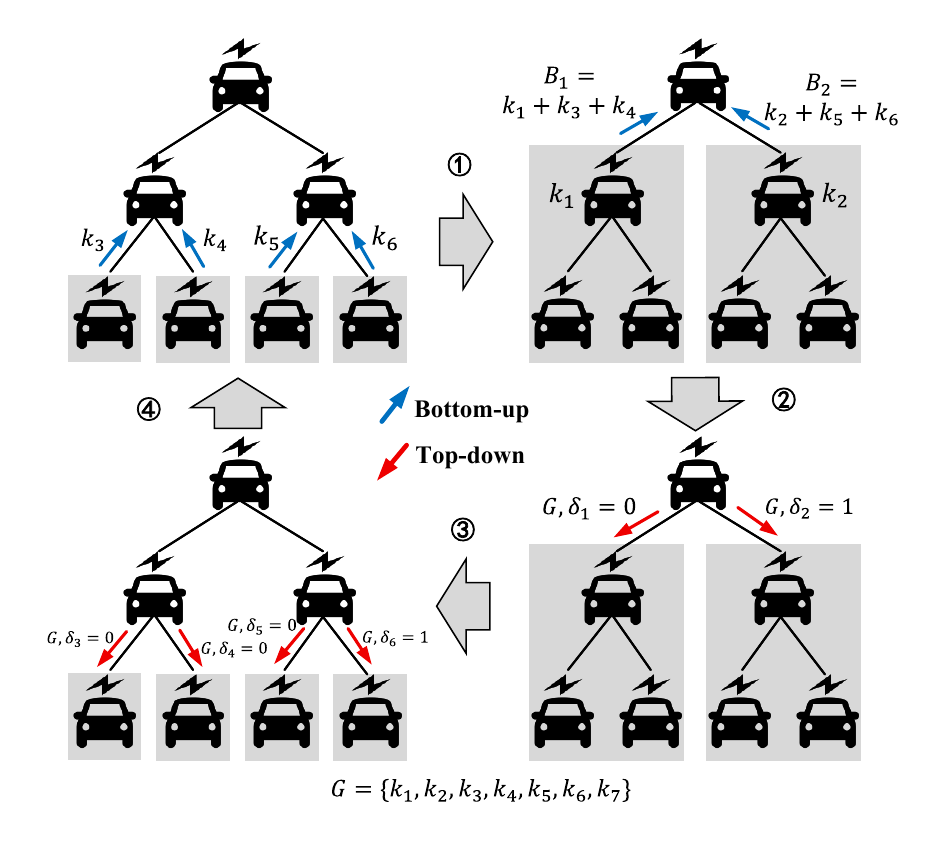}
    \caption{An example of the learning iteration of I-EPOS based on a binary tree topology with $7$ agents (EVs). During the bottom-up phase, each agent sends its charging option $k$ or the branch response $B$ to its parent. During the top-down phase, each agent sends the aggregated options $G$ and decision $\delta$ to its children.}
    \label{fig:tree_structure}
\end{figure}

\subsection{Collective learning for optimized selection}
Decentralized heuristics are required to determine near-optimal, or even highly-performing, solutions for large-scale optimization problems. Thus, this paper uses one such heuristic, the collective-learning algorithm of I-EPOS~\cite{pournaras2018decentralized}, for the coordination of EVs within a time window $t$. 

The approach aims to coordinate EVs, which serve as I-EPOS agents, to minimize system inefficiency while respecting their autonomous behavior for charging selection iteratively. With the generated options, efficient coordinated choices of them are made using a self-organized~\cite{pournaras2020holarchic} tree topology within which agents self-organize their interactions, information exchange and decision-making. At each time window, new EVs establish their proximity-based relationship, e.g., Euclidean distance, to create a new tree topology. After that, they are positioned starting from the leaves up to the root, each interacting with its children and parent~\cite{pournaras2020holarchic}. Fig.~\ref{fig:tree_structure} shows an example of EVs that self-organize into a tree communication structure for information exchange in one iteration. 

The interaction between children and parents of one iteration contains two processes: the bottom-up process and top-down process. During the bottom-up process, each agent $n$ (except for leaf nodes) observes the chosen stations of its children and selects its own option of station. This selection can be formulated as follows:
\begin{equation}
    k^I_{nm} = \underset{k_{nm} \in \mathcal{K}_n}{\text{argmin}} \; f_g (G^I),
    \label{eq:epos}
\end{equation}
\begin{equation}
    f_g (G^I) = \beta_n \cdot \sum_{n \in N_t} \mathcal{O}_1(x_{nm}) + (1 - \beta_n) \cdot \mathcal{O}_2(x_{nm}),
\end{equation}
\begin{equation}
     G^I = \{ k^I_{nm}, \forall n \in N_t \},
\end{equation}
where $k^I_{nm}$ denotes the selected charging option by $n$ at the iteration $I$, $k^I_{nm} = (x_{nm}, t_n, d_n, \mathbb{D}_{nm})$; $f_g$ indicates the overall cost function based on the aggregated options; and $G^I$ is the aggregated options of all agents at the iteration $I$. The agent sums the selected option with the observed options from its children into the branch response, denoted as $B^I_n$ (see the shadow in Fig.~\ref{fig:tree_structure}), and shares this branch response to its parent node (if it is not the root). As a consequence, the root agent can aggregate the selected charging options of all agents and obtain $G^I$. 

During the top-down phase, however, each agent (except leaf nodes) sends the aggregated options of all agents to its children from the root node to leaf nodes such that every agent obtains the aggregated options $G^I$. Apart from that, each agent sends its child a decision $\delta_c$ regarding whether to approve the selected charging option by its child $c$ and its branch in the current iteration, $c \in N_t$. The agent approves if and only if the overall cost defined in Eq.(\ref{eq:obj_overall}) in the current iteration is lower than previous one. The decision of the agent $n$ is formulated as follows:
\begin{equation}
    \underset{\delta_c}{\min} \; f_g [G^{I-1} + \sum_c \delta_c \cdot (B^I_n - B^{I-1}_n)].
\end{equation}
As a result, the child adopts the new selected charging option if approved by its parent, i.e., $\delta_c=1$. Then, it updates the aggregated choices as follows:
\begin{equation}
     G^I = G^{I-1} \backslash k^{I-1}_{nm} \cup k^I_{nm},
\end{equation}
Otherwise, the child keeps the plan selected in the previous iteration, i.e., $\delta_c=0$. Note that the changes of a child $c$ are approved if two conditions hold: (1) The parent agent $n$ approves the changes of its child and its branch with $\delta_c$; (2) The ancestors of the parent approve the changes with $\delta_n$. Thus, the parent agent calculates the decision for its children as $\delta_c = \delta_n \cdot \delta_c$.

Finally, the method converges when all agents no longer reselect new options, i.e., $\delta_c = 1$, $\forall c \in N_t$. EVs execute the selected options of $x_{nm}$, $t_n$ and $d_n$, which are used to update the actual queuing time $\tau^{\text{Q}}_m (t)$ of charging stations through Algorithm~\ref{algorithm1}

\subsection{Charging behavior recommendation} \label{sec:behavior_recommend}
According to the cost function of Eq.(\ref{eq:epos}), the charging behavior $\beta_n$ plays a crucial role in the overall performance of decentralized collective learning. Although $\beta_n$ is private to each EV driver, it is important for system/traffic operators to provide behavior recommendations that balance individual and system-wide objectives defined in Eq.(\ref{eq:objs}): incentivizing more selfish behaviors to minimize driver discomfort or encouraging more altruistic behavior to minimize system inefficiency. For example, if charging slots are scarce relative to requests, neglecting the system inefficiency could result in excessive actual queuing times; in this case, altruistic behavior (i.e., using low $\beta$ to accept higher driver discomfort to reduce system inefficiency) is recommended to substantially shorten queues. In contrast, when charging slots are abundant, EV drivers can behave more selfishly to minimize their discomfort costs without significantly affecting the system.

The recommendation on charging behavior of EVs is formulated as follows:
\begin{equation}
    \beta_n = \gamma \cdot \frac{J^\text{A}_n}{|N_t| (1 + \omega_t)}, \forall n \in N_t
    \label{eq:behavior_recommend}
\end{equation}
where $J^\text{A}_n$ is the number of available charging slots within the range anxiety of request $n$, $\gamma$ is a constant factor, and $\omega_t$ denotes the prediction coefficient to estimate the number of EV charging requests at $t+1$, $0 \leq \omega_t \leq 1$. 

To achieve accurate prediction, \emph{DECharge} leverages the Ordinary Least Squares regression method (OLS) to train these coefficients of $\omega_t$ initially. With the input of historical temporal distribution of EVs, several data from the past time windows $\sum_{t-\Delta t}^t |N_t|$ are selected, and the $N_{t+1}$ is chosen as the target distribution for training. 
Based on this learned model, EVs receive recommendations that consider both current and predicted charging slot availability, as well as the anticipated volume of incoming charging requests. This predictive insight leads to more strategic behavior; for instance, an EV may identify a future shortage of nearby charging slots and proactively adopt a more altruistic strategy at the present (i.e., traveling to distant stations). Such foresight helps alleviate future congestion and contributes to minimizing the overall cost across all time windows.

Note that even though the $\omega_t$ is trained during the initialization process on a third party (e.g., server, cloud), the execution is decentralized, including charging option generation, recommended behavior calculation, and coordination (see Fig.~\ref{fig:framework}). Moreover, the charging behavior is private to EV drivers, and thus they may choose not to follow the recommended behavior by the system, i.e., behaving selfish. Our experiments in Section~\ref{sec:eval_differ} will analyze the influence of this adversary behavior.

\begin{figure}[!t]
    \centering
    \includegraphics[width=0.7\linewidth]{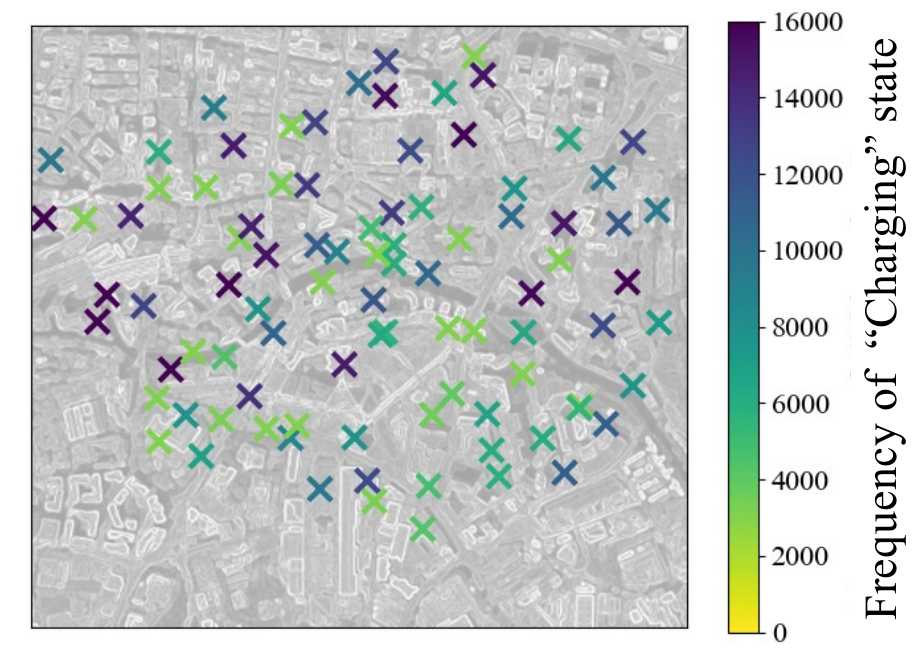}
    \caption{Spatial distribution of charging stations in Paris and their frequency of ``Charging'' state (i.e., the number of times each charging station has been requested by EVs over a year).}
    \label{fig:station_dist}
\end{figure}

\begin{figure}[!t]
    \centering
    \subfigure[Hour of day.]{
        \includegraphics[width=0.45\linewidth]{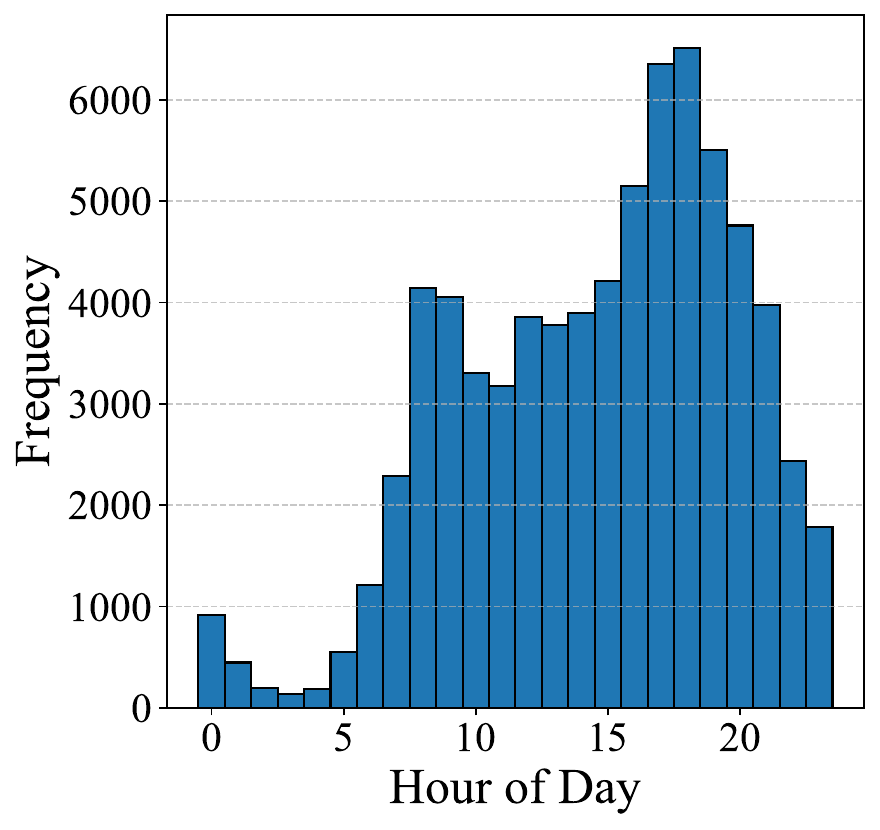}
        \label{fig:frequency_time}
	}
    \subfigure[Demanded charging time.]{
        \includegraphics[width=0.45\linewidth]{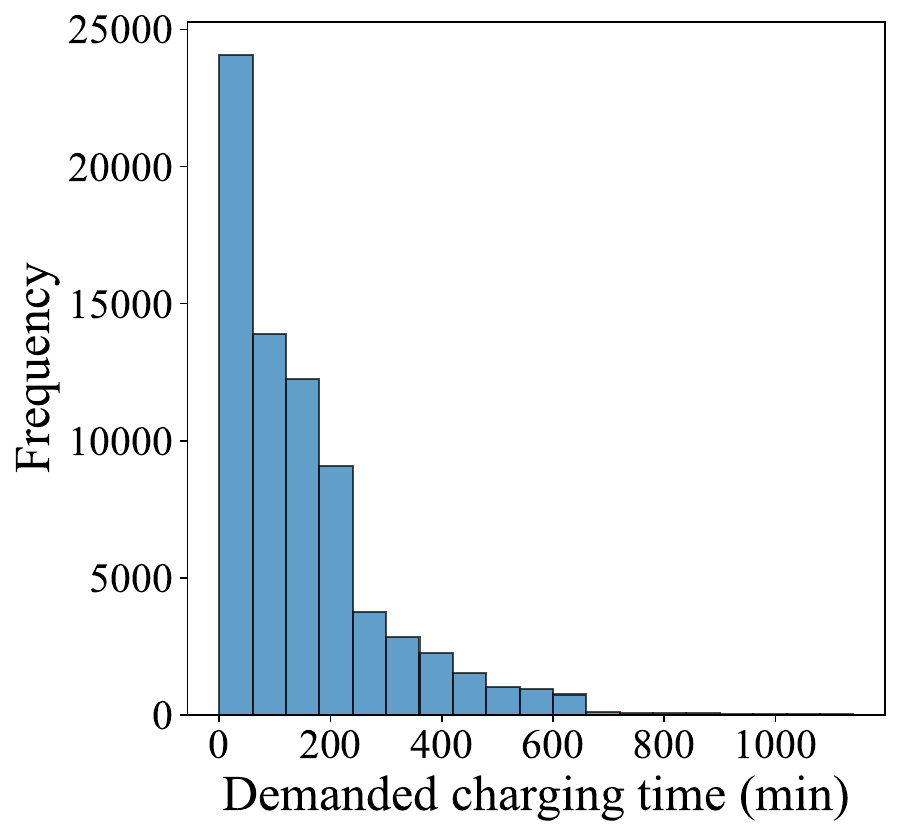}
        \label{fig:frequency_duration}
	}
    \caption{Temporal distribution of EV charging requests and their demanded charging time.}
    \label{fig:frequency}
\end{figure}

\section{Experimental Methodology} \label{sec:settings}
This section firstly introduces the experimental settings, including the simulation based on real-world datasets, experimental scenarios and algorithms, for obtaining valid and reproducible results. The metrics and baselines used for the performance evaluation are also introduced.

\subsection{Dataset and implementation details}
We simulate EV requests based on real station activity and temporal demand data of EVs to preserve the realism of actual charging behavior. While existing datasets capture real EV charging requests, they typically reflect historical, uncoordinated behaviors, which do not represent how EVs would behave and make choices of charging stations under a decentralized coordinated mechanism~\cite{liu2020incorporating,zhang2023charging}. Moreover, due to privacy constraints, they often lack detailed driver information such as geographical coordinates~\cite{national2025data,kujala2018collection}. By synthesizing EV requests based on real station activity patterns and empirical temporal demand distribution of EVs, we retain the realism of actual infrastructure usage while conducting controlled and repeatable experiments.
  
We use the ENS Challenge Data-EV charging Stations Usage~\cite{planete2025}, a dataset that records the locations and state of $91$ charging stations in Paris for one year. Every charging station in the dataset has three charging slots (i.e., $273$ charging slots in total), each with a state of ``Charging'', ``Available'' or ``Down'' depending on the environmental factors such as weather and traffic conditions. As shown in Fig.~\ref{fig:station_dist}, the frequency of ``Charging'' state indicates the number of times each charging station has been requested by EVs over a year.
 
The temporal and spatial distribution of simulated EV charging requests are derived from real-world data to ensure realistic but controllable experimental conditions. The temporal distribution is obtained from the Multi-faceted Analysis of Electric Vehicle Charging Transactions dataset~\cite{baek2024dataset}, which records the charging start and end times, energy demands and charging efficiency of 2,238 EVs over one year in South Korea, averaging about 180 charging requests per day. As shown in Fig.~\ref{fig:frequency_time} and \ref{fig:frequency_duration}, most EVs request charging for less than one hour, with peak demand occurring between 16:00 and 20:00, typically after work hours. This South Korea dataset does not include the spatial coordinates of EVs and is used solely simulate temporal charging requests. Thus, we calibrate the data to Paris through two methods: (1) Model the temporal distribution of EVs and their demanded charging time using kernel density estimation; (2) Assign each simulated EV to the location of a real charging station in Paris, i.e., the same coordinates. Specifically, the number of EVs at each station is proportional to its historical charging frequency~\cite{planete2025}, meaning that stations with higher past utilization (i.e., ``Charging'' state frequency) generate more simulated requests. For example, stations located in the western part of the city exhibit higher activity (see Fig.~\ref{fig:station_dist}) and thus have a higher number of simulated EV charging requests.

\subsection{Experimental scenarios and algorithms}
The experimental scenarios include the \textit{basic charging scenario} and \textit{complex charging scenario}. The basic synthetic scenario divides a day into 12 time windows with a length of 120 min. Half of the charging slots are assumed to be available, and hence we set the ratio of available charging slots as $0.5$. The estimated travel distance is calculated by the Euclidean distance between the positions of EVs and stations. The constant parameters in the formulation of driver discomfort are set $\alpha_1 = \alpha_2 = 1$. In the \textit{complex charging scenario}, however, the parameter settings are varied, including the ratio of available charging slots and the number of time windows.

For the coordinated optimized selection using I-EPOS\footnote{EPOS is available at: https://github.com/epournaras/EPOS.}, 
the collective learning agents (i.e., EV charging requests in a day) are self-organized in a balanced binary tree communication structure. Note that agents determine their parents and children according to their relative distance, they should be repositioned to test robustness and adaptability of the algorithm to different spatial configurations of EVs. Therefore, the algorithm repeats 40 times by changing the random position of the agents in the tree\footnote{More information about the influence of the tree topology and agents' positioning on the tree is illustrated in earlier work~\cite{pournaras2020holarchic}.}. Each repetition of I-EPOS consists of 40 learning iterations, alternating between bottom-up and top-down. For the validation of the proposed algorithm, we run I-EPOS for all 365 data records, each representing the EV charging requests within a day.

\subsection{Evaluation metrics}
Apart from the abstract metrics such as driver discomfort and system inefficiency defined in Eq.(\ref{eq:obj_discomfort}) and Eq.(\ref{eq:obj_inefficiency}) respectively, there are five more metrics that evaluate the performance of our approach and baseline methods. The metrics are detailed as follows: (1) \textbf{Overall operational cost:} It is the average value of both driver discomfort and system inefficiency. (2) \textbf{Relative travel distance:} It measures the estimated travel distance (km) to charging stations per EV charging request on average: $\sum_{n=1}^N \sum_{m=1}^M D_{nm} / N$. (3) \textbf{Actual queuing time:} It computes the queuing time (hour) per EV on average after it is assigned to a charging station: $\sum_{t=1}^T \sum_{m=1}^M \tau^{\text{Q}}_{m} (t) / N$. (4) \textbf{Estimated waiting time:} It is the summation of both actual queuing time and travel time, where the travel time is calculated via the relative travel distance divided by the speed of EV under moderated traffic (set as 30 km/h). (5) \textbf{Station demand:} It calculates the accumulated energy demand (kJ) of EVs on each station over a day using charging efficiency derived from the EV charging transaction dataset~\cite{baek2024dataset}.

\subsection{Baselines}
A direct and fair comparison between the proposed approach based on the collective learning of I-EPOS, and existing methods is challenging due to the scarcity of relevant algorithms that address the same problem setting of decentralized EV charging control. As outlined in Section~\ref{sec:related}, conventional distributed optimization techniques (e.g., particle swarm optimization and genetic algorithms) are not readily applicable to the multi-objective discrete-choice combinatorial optimization problem defined in Section~\ref{sec:model}, particularly under the constraint of decentralized execution. 

For this reason, we use the state-of-the-art baselines of \textit{COHDA}~\cite{hinrichs2014cohda} (\textit{Combinatorial Optimization Heuristic for Distributed Agents}) and \textit{MGM} (\textit{Maximum Gain Message})~\cite{acs2024distributed} for comparison. Both are decentralized coordination methods where each agent iteratively decides its action by calculating the potential gain in the overall objective and sharing it with neighbors. Agents with the highest gain are allowed to change their actions, while others wait, ensuring conflicts are avoided. Since agents have full view of the system by sharing full information with others, these methods have higher communication or computational overhead than I-EPOS that focuses on hierarchical acyclic graphs such as self-organized trees to perform a cost-effective decision-making and aggregation of choices. Table~\ref{tab:compare} compares the complexity of these approaches in both agent- and system-level. For fair comparison, we limit the information exchange of baseline methods to prove the cost-effectiveness of \emph{DECharge}. 

Furthermore, given the limited number of relevant decentralized algorithms, we focus our evaluation on a set of ablation studies, each designed to isolate and assess the contribution of key components. More baseline methods are listed as follows: (1) \emph{Greedy:} It indicates that EV drivers (i.e., charging requests) always behave selfish without coordination by I-EPOS. They choose the charging option with the minimum driver discomfort. (2) \emph{Distance-Only Coordination (DOC):} It defines that the driver discomfort $\mathbb{D}_{nm}$ only include travel distance $D_{nm}$, excluding observed queuing time $Q_{nm}$, i.e., $\mathbb{D}_{nm} = D_{nm}$. (3) \emph{Static Initial Coordination (SIC):} It defines a single time window ($T = 1$) during which all EVs in a day make their charging station selection at the same time. In this setting, the observed queuing time $Q_{nm}$ of EVs is 0.

\begin{table}[!t]
\centering
\caption{Complexity comparison of decentralized approaches for discrete-choice combinatorial optimization problem.}
\centering
\label{tab:compare}
\resizebox{\linewidth}{!}
{
    \begin{tabular}{llll}
    \hline
    Approaches & Computational cost & Communication cost \\ \hline
    \textit{I-EPOS}~\cite{pournaras2018decentralized} &
    \tabincell{l}{agent: $O(KI)$; \\ system: $O(KI\ log\ N)$}
    &
    \tabincell{l}{agent: $O(I)$; \\ system: $O(I\ log\ N)$} \\

    \textit{COHDA}~\cite{hinrichs2014cohda} &
    \tabincell{l}{agent: $O(KI)$; \\ system: $O(KI)$}
    & 
    \tabincell{l}{agent: $O(IN)$; \\ system: $O(IN)$} \\ 

    \textit{MGM}~\cite{acs2024distributed} &
    \tabincell{l}{agent: $O(KI)$; \\ system: $O(KIN)$}
    &
    \tabincell{l}{agent: $O(IN)$; \\ system: $O(IN)$} \\ \hline
    
    \multicolumn{4}{p{7.5cm}}{ $K$: number of options, $I$: number of iterations, $N$: number of agents} \\
    \end{tabular}
}
\end{table}

\begin{table*}[!t]  
	\caption{Performance comparison of different approaches on the basic charging scenario (Lower-is-Better).}  
	\centering
    \resizebox{\linewidth}{!}{
    	\begin{tabular}{lcccccc}  
		\hline  
		Metrics & \emph{COHDA} & \emph{MGM} & \emph{Greedy} & \emph{DOC} & \emph{SIC} & \emph{DECharge} \\
		\hline    
		Driver discomfort    & $0.106 \pm 0.094$ & $1.227 \pm 0.102$ & $0.070 \pm 0.068$ & $0.440 \pm 0.375$ & $0.208 \pm 0.021$ & $0.094 \pm 0.066$  \\
		System inefficiency  & $1.871 \pm 0.347$ & $1.592 \pm 0.440$ & $2.222 \pm 0.434$ & $1.586 \pm 0.276$ & $7.364 \pm 1.133$ & $1.550 \pm 0.279$  \\
        Overall operational cost  & $0.989 \pm 0.235$ & $1.410 \pm 0.271$ & $1.146 \pm 0.251$ & $1.013 \pm 0.326$ & $3.786 \pm 0.577$ & $\textbf{0.822} \pm \textbf{0.172}$  \\
		Relative travel distance (km)      
        & $0.204 \pm 0.033$ & $4.608 \pm 0.065$ & $0.162 \pm 0.031$ & $0.171 \pm 0.030$ & $1.039 \pm 0.103$ & $0.276 \pm 0.040$  \\
		Actual queuing time (h)       
        & $0.156 \pm 0.075$ & $0.305 \pm 0.202$ & $0.309 \pm 0.178$ & $0.109 \pm 0.057$ & $0.295 \pm 0.097$ & $0.006 \pm 0.005$  \\
        Estimated waiting time (h)
        & $0.163 \pm 0.076$ & $0.459 \pm 0.204$ & $0.314 \pm 0.179$ & $0.115 \pm 0.058$ & $0.330 \pm 0.100$ & $\textbf{0.015} \pm \textbf{0.006}$  \\
        Max station demand (kJ)   
        & $180.3 \pm 38.37$  & $261.6 \pm 27.55 $ & $283.8 \pm 48.41$  & $167.9 \pm 37.57$ & $153.37 \pm 20.98$ & $156.6 \pm 35.10$  \\
		\hline  
	\end{tabular}  
    }
	\label{table:basic}
\end{table*}

\section{Performance Evaluation} \label{sec:evaluation}
This section assesses the different system parameters and illustrates the results of the evaluated scenarios of both basic charging and complex charging. 

\subsection{Overall performance}
We first evaluate the performance metrics of different approaches in the basic charging scenario with $12$ time windows and $50\%$ of available charging slots. As shown in Table~\ref{table:basic}, the proposed \emph{DECharge} achieves lower overall operational cost and shorter estimated waiting time compared to all baseline methods. Relative to \emph{COHDA}, \emph{DECharge} reduces driver discomfort and system inefficiency with the same level of information exchange, achieving a $16.89\%$ reduction in overall operational cost and demonstrating its cost-effectiveness. Moreover, \emph{DECharge} substantially reduces actual queuing time, resulting in an average estimated waiting time reduction of around $8.9$ minutes per EV compared to \emph{COHDA}, and achieving the reduction of $26.64$ minutes per EV compared to \emph{MGM}. 

\begin{figure}[!t]
    \centering
    \includegraphics[width=0.85\linewidth]{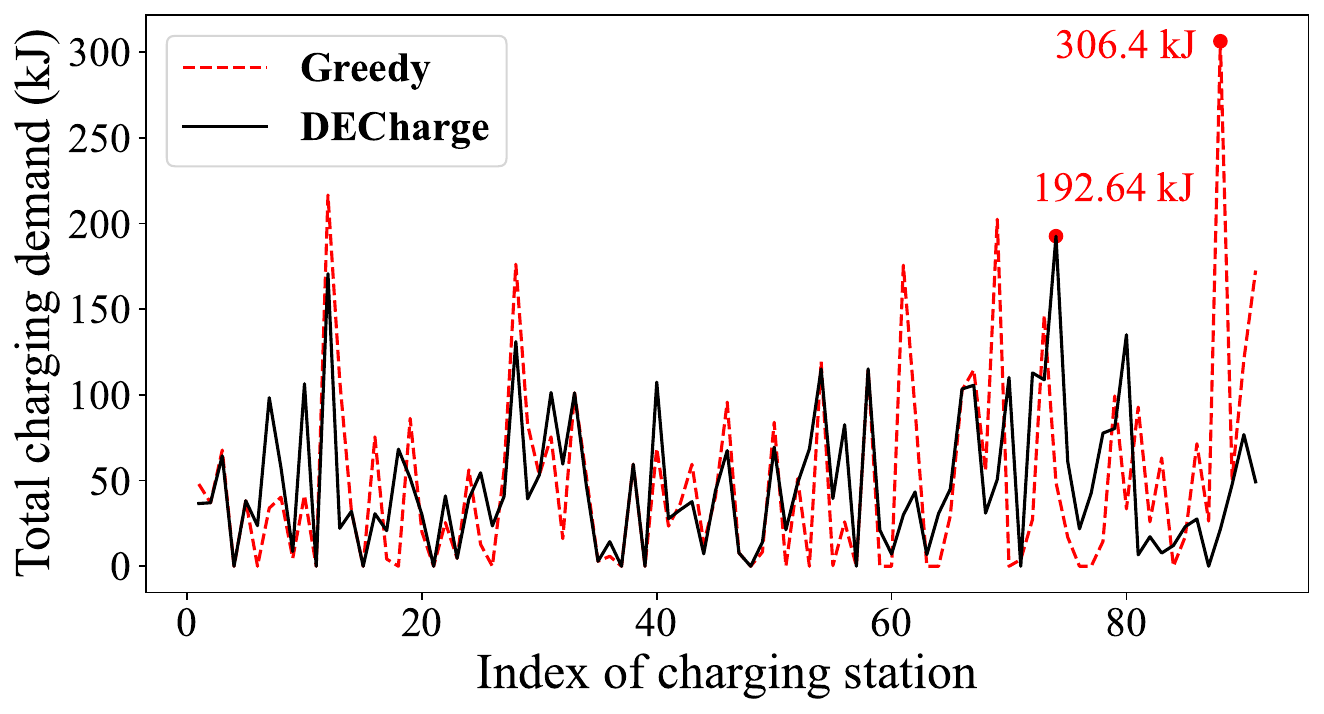}
    \caption{Distribution of station demand across different approaches in the basic charging scenario.}
    \label{fig:charging_demand}
\end{figure}

The ablation studies reveal further insights on the effectiveness of I-EPOS and driver discomfort formulation in Eq.(\ref{eq:obj_discomfort}). \emph{Greedy} minimizes driver discomfort and relative travel distance among all approaches, but at the cost of high system inefficiency due to lack of coordination. As a result, its actual queuing time is roughly $18.2$ minutes per EV longer than \emph{DECharge}, and its station demand distribution is more imbalanced, with a higher peak station demand (see Fig.~\ref{fig:charging_demand}). When driver discomfort is defined solely by travel distance (\emph{DOC}), average travel time decreases by $0.21$ minutes compared to \emph{DECharge}, but actual queuing time increases by $6.12$ minutes per EV. In the static initial coordination setting (\emph{SIC}), the charging imbalance is mitigated by lowering maximum station demand, but both driver discomfort and system inefficiency increase sharply, leading to an estimated waiting time that is $18.9$ minutes per EV longer than \emph{DECharge}.

\begin{figure}[!t]
    \centering
    \includegraphics[width=\linewidth]{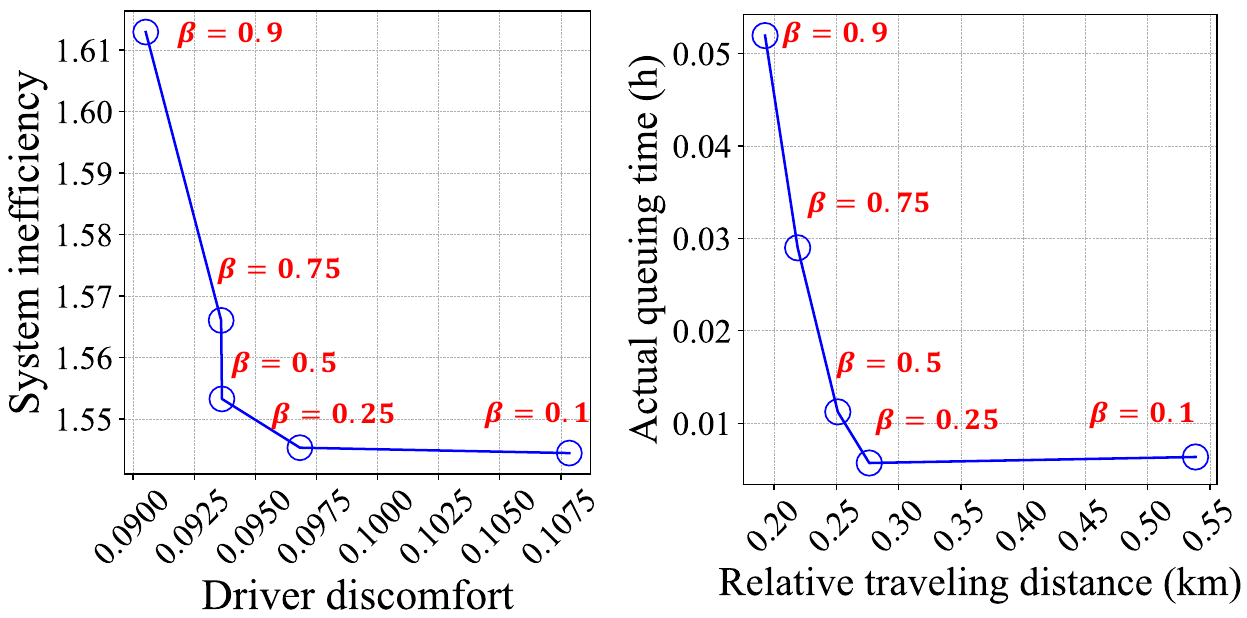}
    \caption{Performance comparison of \emph{DECharge} by changing charging behavior $\beta$.}
    \label{fig:beta_pareto}
\end{figure}

Fig.~\ref{fig:beta_pareto} illustrates the effect of different charging behavior $\beta$. As $\beta$ increases, both driver discomfort and relative travel distance decrease, while both system inefficiency and actual queuing time increase. Combined with the results in Table~\ref{table:basic}, Fig.~\ref{fig:beta_pareto} demonstrates that EVs are recommended with optimal $\beta$ from $0.25$ to $0.5$, achieving the Pareto optimality. 

Overall, jointly minimizing two objectives of driver discomfort and system inefficiency effectively reduces both travel distance and queuing time for EVs requesting charging, while reducing imbalanced station energy demand. The overall operational cost is a representative metric for evaluation as it is approximately proportional to estimated waiting time. Compared to baseline methods, \emph{DECharge} delivers highly cost-effective performance by integrating coordinated decision-making with timely queuing-time observations. The real-time observation on queuing time makes the learning efficient to find the optimized solution as it mitigates the conflicts of these two objectives. The experimental results also support the theoretical properties defined in Appendix~\ref{sec:theoretical}.

\begin{figure*}[!t]
    \centering
    \includegraphics[width=\linewidth]{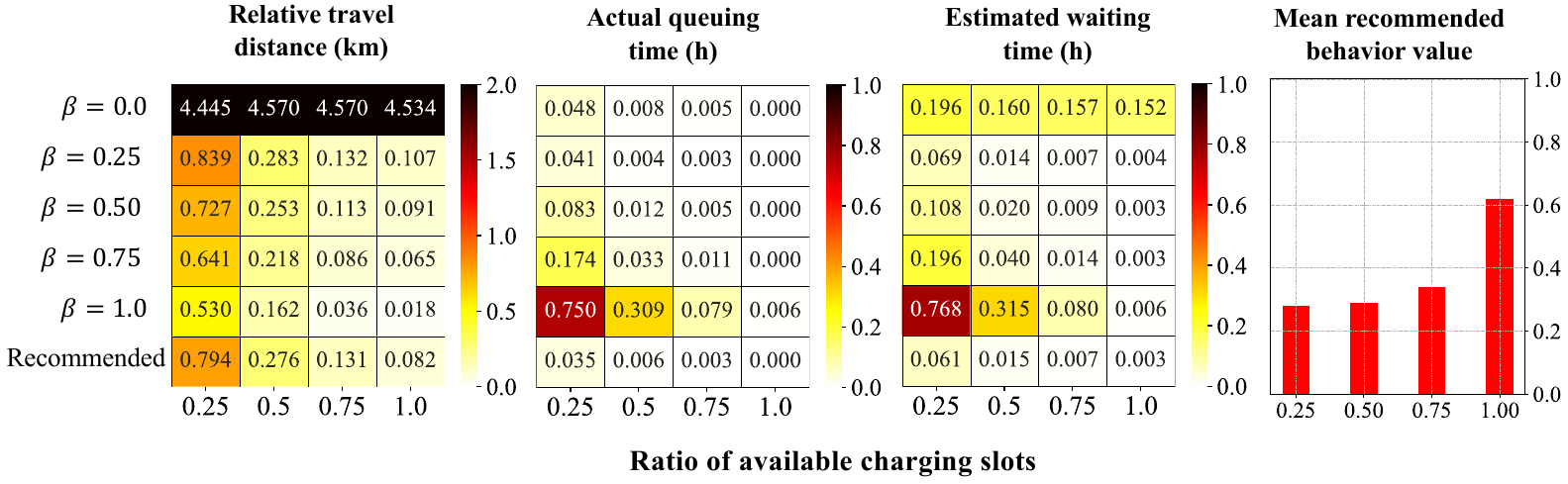}
    \caption{The performance comparison of \emph{DECharge} over different charging behavior and the ratio of available charging slots. The ``Recommended'' refers to  \emph{DECharge} using charging behavior recommendation. All behavior values of EVs recommended by the system is summed up to obtain the mean recommended behavior value per EV.}
    \label{fig:heatmap_complex}
\end{figure*}

\subsection{Effect of different parameters} \label{sec:eval_differ}
To study the complex charging scenario, we vary two key factors: (1) the ratio of available charging slots, from $0.25$ to $1.0$, (2) the number of time windows, from $2$ to $24$, and (3) the percentage of EV drivers that behave selfishly, from $0\%$ to $40\%$. In addition, to evaluate the charging behavior recommendation in the proposed \emph{DECharge}, we test a variant where all EVs adopt the same fixed behavior value $\beta = \beta_n$, $\forall n \leq N$. The range of $\beta$ is set from $0$ to $1$.

We randomly select a number of charging slots in all stations to be ``Available'' and others are ``Down'', which simulates the unexpected station outages in real world. As shown in Fig.~\ref{fig:heatmap_complex}, when the ratio of available charging slots decreases, the number of EV charging requests significantly exceeds available capacity (i.e., high request density), resulting in longer actual queuing time. If EVs are recommended to choose lower $\beta$ to behave more altruistic instead, e.g., reducing $\beta$ from $1.0$ to $0.25$, they decrease actual queuing time by $84.47\%$ and lowers estimated waiting time by $37.58\%$, albeit at the cost of increasing relative travel distance. Note that setting $\beta = 0.0$ is undesirable, as ignoring driver discomfort entirely leads to higher travel distance and longer waiting time. In contrast, when charging slots are sufficient, EV drivers can adopt more selfish behaviors (e.g., $\beta \in [0.75, 1.0]$), as minimizing the actual queuing time, which is already incorporated into driver discomfort, requires little coordination. 

The charging behavior recommendation mechanism in \emph{DECharge} adaptively determines the optimal $\beta$ at each time window based on the predicted ratio of charging requests to available slots, defined in Eq.(\ref{eq:behavior_recommend}), achieving the minimum estimated waiting time, see Fig.\ref{fig:heatmap_complex}. It further validates that EVs adopting adaptive charging behaviors outperform those following fixed moderate behaviors (i.e., $\beta = 0.5$). Similarly, changing the number of EVs in a day but fixing the ratio of available charging slots also varies the request density, which is equivalent to change the availability of charging slots only. Therefore, the results demonstrate the adaptability and resilience of the proposed approach under complex EV distribution and even high fractions of station failures. 

\begin{figure}[!t]
    \centering
    \subfigure{
        \includegraphics[width=0.45\linewidth]{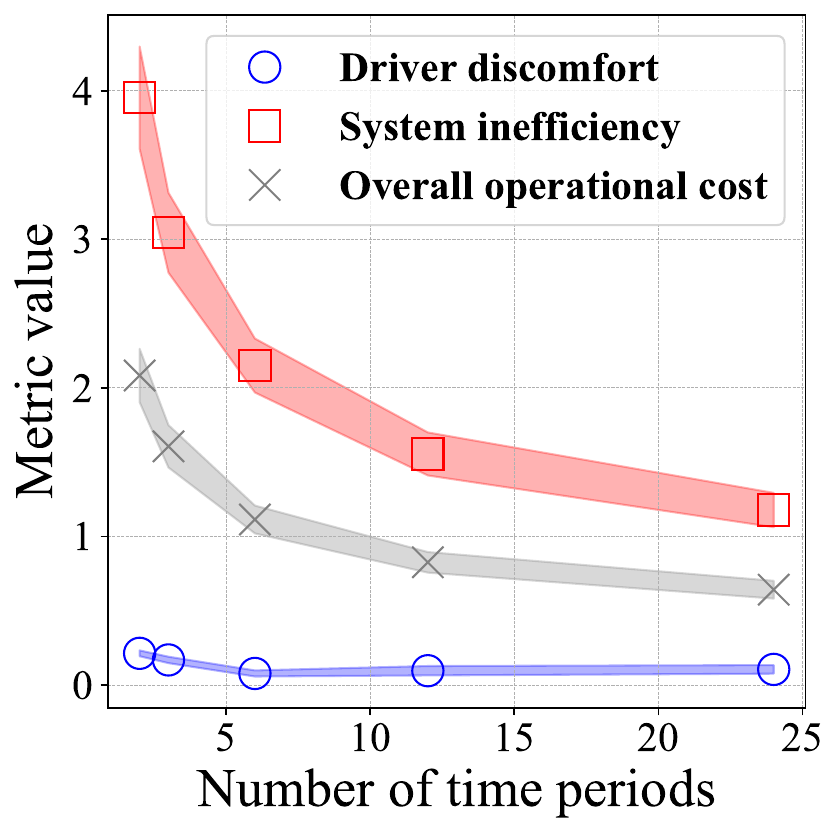}
        \label{fig:time_metric1}
	}
    \subfigure{
        \includegraphics[width=0.48\linewidth]{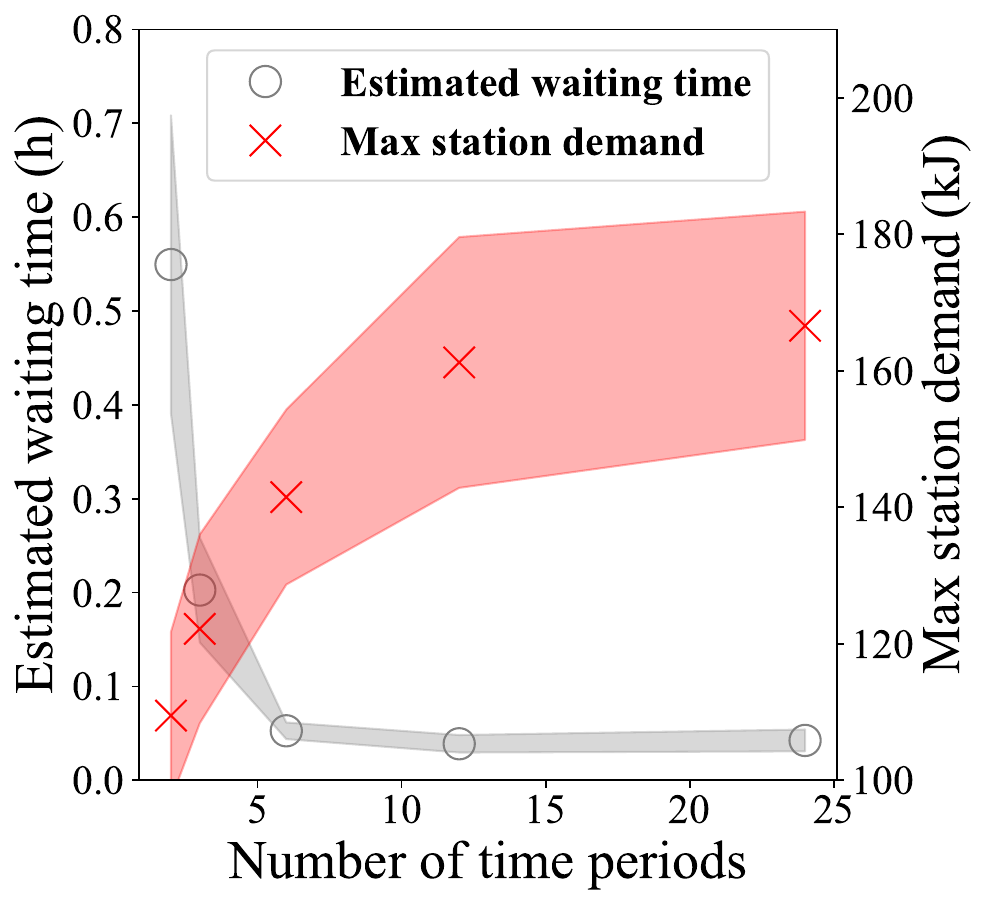}
        \label{fig:time_metric2}
	}
    \caption{The performance comparison of \emph{DECharge} over different number of time windows. The shadow represents the error.}
    \label{fig:time}
\end{figure}

When increasing the number of time windows from 2 to 24 (Fig.~\ref{fig:time}), the number of coordination instances between EV charging requests grows, while the window length for observing queuing time per EV shortens from 12 hours to 1 hour. More frequent and timely observations reduce both driver discomfort and system inefficiency, which converge after approximately 12 time windows. As a consequence, estimated waiting time decreases by $92.28\%$, although the maximum station demand across all stations rises by $52.16\%$. This confirms the effectiveness of real-time queuing-time observations, achieved at the expense of a less balanced energy demand across stations.

\begin{figure}[!t]
    \centering
    \subfigure{
        \includegraphics[width=0.46\linewidth]{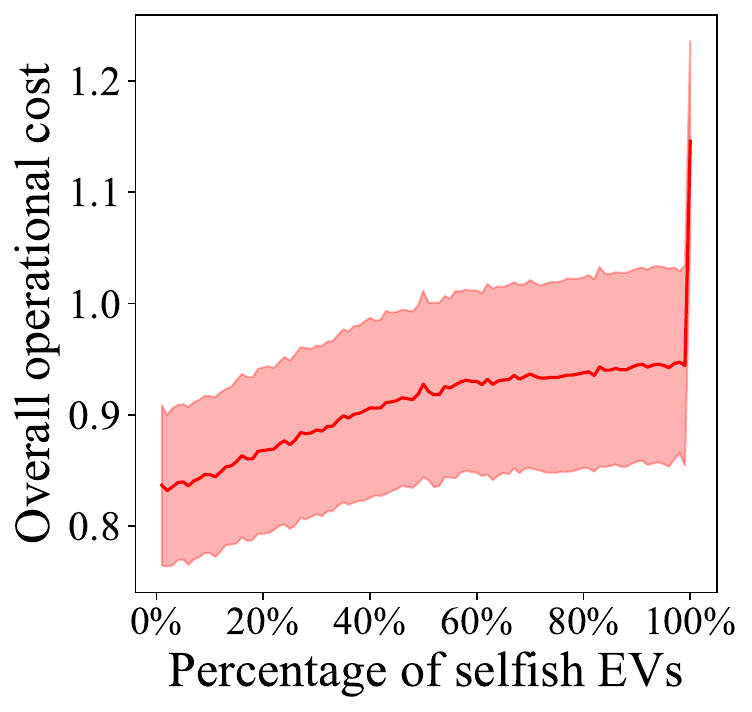}
        \label{fig:misbehaved1}
	}
    \subfigure{
        \includegraphics[width=0.46\linewidth]{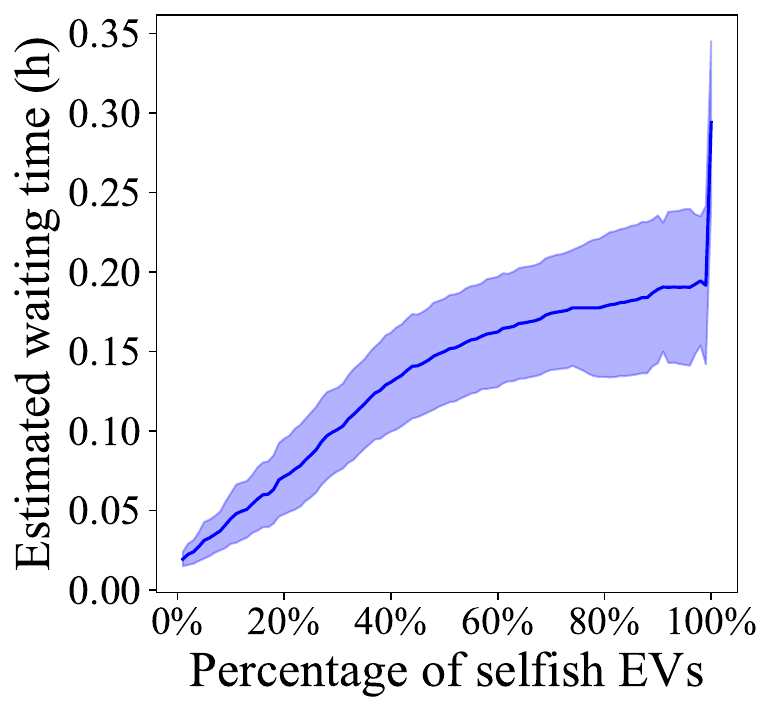}
        \label{fig:misbehaved2}
	}
    \caption{The performance comparison of \emph{DECharge} over different percentages of EVs that behave selfish $\beta = 1$.}
    \label{fig:misbehaved}
\end{figure}

Since charging behaviors are only recommendations, EVs in the decentralized charging infrastructure remain free to deviate from them. For example, some EV drivers may still act selfishly without coordinating with others. Figure~\ref{fig:misbehaved} evaluates the performance of \emph{DECharge} under such adversary behavior. We randomly select a number of EV charging requests in a day to behave selfishly ($\beta = 1.0$). As the percentage of selfish EV drivers increases, more vehicles choose charging options with the lowest driver discomfort. When the percentage increases from $0\%$ to $99\%$, the overall operational cost increases by $12.82\%$, resulting in a $10$ minutes rise in estimated waiting time per EV. However, when the percentage increases to $100\%$, which means all EVs act selfishly without coordination (same as \emph{Greedy}), the estimated waiting time surges. The results illustrate that \emph{DECharge} coordinates EVs efficiently even though $99\%$ of EVs deviate from the behavior recommendations, which validates the high resilience of the proposed approach.

\subsection{Discussion and new insights}
In summary, the experimental results validate the superior performance of the proposed \emph{DECharge}. Several scientific insights on \emph{DECharge} are listed below.

\cparagraph{Achieving EV charging system efficiency and reliability}
\emph{DECharge} efficiently coordinates EVs to minimize system inefficiency while making a Pareto-optimal trade-off between their travel distance and queuing time, leading to $97.72\%$ lower charging waiting time than the approach without coordination. Meanwhile, it ensures a balanced distribution of energy demand across stations, reducing the energy demand peak by $37.13\%$ (i.e., prevention of power peak) and improving the overall reliability of the charging infrastructure. The choice of I-EPOS further improves the cost-effectiveness of the coordination by $29.30\%$ (i.e., overall operational cost) compared to other distributed optimization methods.

\cparagraph{Real-time decision-making through privacy-preserving driver discomfort modeling}
\emph{DECharge} divides the day into multiple time windows, allowing EVs requesting charging within each time window to dynamically adjust their station selection based on the current queuing times across all stations. This real-time observation, incorporated into the driver discomfort formulation in Eq.(\ref{eq:obj_discomfort}), effectively reduces several hours of waiting time per EV. Furthermore, the driver discomfort is designed to preserve sensitive information of EVs, such as location, state of charge and range anxiety. 

\cparagraph{Adaptive charging behavior recommendation for dynamic environments}
The proposed charging behavior recommendation by system/traffic operators guides EVs to adjust their charging strategies in response to complex, dynamic environments, such as variations in station capacity and the spatio-temporal distribution of EVs. As shown in Fig.~\ref{fig:heatmap_complex}, when charging request density is high, EVs are encouraged to act altruistically by selecting options with higher discomfort, e.g., choosing a distant station; when density is low, they are admitted to prioritize lower discomfort and select the nearest stations. This adaptability minimizes the waiting time of EVs under diverse operating conditions, outperforming the scenarios where all EVs choose moderate behaviors.

\cparagraph{Resilience to adversary behavior and system trustworthiness}
The charging infrastructure remains robust unless $100\%$ of EVs deviate from recommended behaviors, with minimal impact on cost and waiting time. Since selfish EVs benefit from following recommendations, i.e., each can save approximate $18$ minutes waiting time, they are naturally incentivized to cooperate. This resilience supports stable performance and greater trust in coordinated charging, benefiting broader transportation networks.

\section{Conclusion and Future Work} \label{sec:conclusion}
In conclusion, this paper presents \emph{DECharge}, a decentralized collective learning-based coordination framework that redefines how EVs autonomously select charging stations under uncertainty. By balancing driver discomfort and system inefficiency through adaptive charging behaviors, \emph{DECharge} allows EVs to dynamically choose close or distant stations based on real-time station capacities and spatio-temporal distribution of charging requests. The framework achieves a Pareto-optimal trade-off between travel distance and queuing time, resulting in significantly lower waiting time compared to baseline methods and EVs maintaining moderate behaviors. Moreover, \emph{DECharge} maintains strong performance even under adverse conditions, such as station failures or the presence of adversarial EVs. These experimental results highlight the potential of \emph{DECharge} to enhance scalability, resilience and trustworthiness of decentralized EV charging infrastructure.

Apart from that, several directions remain for further exploration: (1) Investigate how practical aspects such as road network topology, types of charging locations (e.g., residential, industrial, public), and weather conditions affect travel/arrival time, charging behavior and station selection. (2) Extend the proposed solution with reinforcement learning to generate long-term behavior recommendations that improve coordination over extended horizons.


%

\section*{Acknowledgment}
{
This paper is supported by the European Union’s HORIZON Research and Innovation Programme under grant agreement No 101120657, project ENFIELD (European Lighthouse to Manifest Trustworthy and Green AI). This paper is also supported by a UKRI Future Leaders Fellowship (MR-/W009560/1): 
\emph{Digitally Assisted Collective Governance of Smart City Commons–ARTIO}.
}

\bibliographystyle{unsrt}
\bibliography{reference}

\appendices

\section{Theoretical analysis} \label{sec:theoretical}
This appendix analyzes the theoretical properties of our formulated model, focusing on the objective functions defined in Eq.(\ref{eq:obj_discomfort}) and Eq.(\ref{eq:obj_inefficiency}) as well as their mathematical proof.

\textbf{Property 1}. \textit{For imbalanced distribution of EV charging requests within a time window $t$, the objective to minimize their driver discomfort is inverse proportional to the objective to minimize system inefficiency.}

\begin{figure}[!t]
    \centering
    \includegraphics[width=0.9\linewidth]{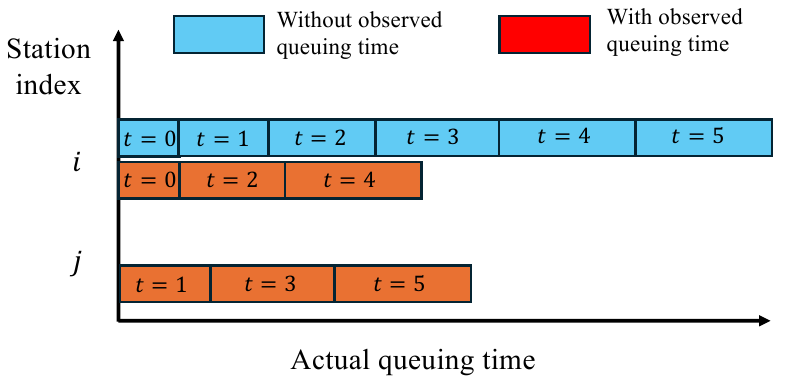}
    \caption{The update of actual queuing times of station $i$ and $j$ within $6$ time windows. The results are also compared with the method that the observed queuing time is not considered in driver discomfort.}
    \label{fig:appendix_prove}
\end{figure}

\begin{proof}
To simplify the proof, we assume each charging station has only one slot $J=1$, and the constant parameter are set $\alpha_1 = \alpha_2 = 1$. The spatial distribution of EVs is based on a Gaussian model, i.e., the number of EVs located at the center of the map is higher than that of margins. The objective function for EVs within $N_t$ are reformulated as follows:
\begin{equation}
    \mathcal{O}'_1 (x_{nm}) = \sum_{m=1}^M [D_{nm} + \tau^{\text{Q}}_m(t)] \cdot x_{nm},
\end{equation}
\begin{equation}
    \mathcal{O}'_2 (x_{nm}) = \sqrt{\frac{1}{M} \sum_{m=1}^M [\tau^{\text{Q}}_m(t) + \sum_{n \in N_t} d_n \cdot x_{nm}]^2},
\end{equation}
where the actual queuing time $\tau^{\text{Q}}_m(t)$ of station $m$ at time window $t$ can be formulated as below:
\begin{equation}
    \tau^{\text{Q}}_m(t) = \sum_{t_0 \leq t} \sum_{n \in N_{t_0}} d_n \cdot x_{nm} - t.
    \label{eq:appendix_queue}
\end{equation}
Assume that there exists a solution $x^*_{nm} \in \{0, 1\}$, $n \leq N$, $m \leq M$ such that both objectives are minimized simultaneously $x^*_{nm} = \arg \min \mathcal{O}'_1 (x_{nm}) = \arg \min \mathcal{O}'_2 (x_{nm})$. 

Given $x^*_{nm} = \arg \min \mathcal{O}'_1 (x_{nm})$, when $t=0$ and actual queuing time $\tau^{\text{Q}}_m(0) = 0$, EVs only choose the nearest charging stations. Due to the imbalanced distribution of EV charging requests, some stations are chosen more frequently than others. For example, the station $i$ in the traffic-congested areas is selected more than that $j$ in non-congested areas, $i,j \leq M$, i.e., $\sum_{n \in N_t} d_n \cdot x_{ni} > \sum_{n \in N_t} d_n \cdot x_{nj} = 0$. This skewed allocation caused by min driver discomfort $\mathcal{O}'_1$ results in higher system inefficiency $\mathcal{O}'_2$, and $x^*_{nm} \neq \arg \min \mathcal{O}'_2 (x_{nm})$. 

When $t>0$, EVs observe the higher queuing time $\tau^{\text{Q}}_m (t)$ of station $i$ in traffic-congested area compared to the station $j$ in non-congested areas. It means the driver discomfort of choosing station $i$ is higher than that of choosing station $j$, i.e., $D_{ni} + \tau^{\text{Q}}_m(t) > D_{nj} + \tau^{\text{Q}}_m(t)$. Thus, they all select the station $j$ and leads to imbalanced charging demand on stations, i.e., $\sum_{n \in N_t} d_n \cdot x_{nj} > \sum_{n \in N_t} d_n \cdot x_{ni} = 0$. If the time window $t$ increases, according to Eq.(\ref{eq:appendix_queue}), the actual queuing time of station $i$ and $j$ is updated iteratively, see Fig.~\ref{fig:appendix_prove}. This results in imbalanced actual queuing time across stations, i.e., $\tau^{\text{Q}}_i(t) + \sum_{n \in N_t} d_n \cdot x_{ni} \neq \tau^{\text{Q}}_j(t) + \sum_{n \in N_t} d_n \cdot x_{nj}$. Therefore, the system inefficiency is increased caused by decreasing driver discomfort and $x^*_{nm} \neq \arg \min \mathcal{O}'_2 (x_{nm})$.

In contrast, given $x^*_{nm} = \arg \min \mathcal{O}'_2 (x_{nm})$, EVs are evenly allocated to stations. EVs are required to travel to distant stations, i.e., $D_{nm}$ increases, $m \leq M$, even though the actual queuing time of stations $\tau^{\text{Q}}_m(t)$ are the same. Hence, the driver discomfort of EVs $D_{nm} + \tau^{\text{Q}}_m(t))$ increases, and $x^*_{nm} \neq \arg \min \mathcal{O}'_1 (x_{nm})$. As a consequence, this contradicts the assumption that both objectives are simultaneously minimized by the same $x^*_{nm}$. The two objectives are inverse proportional to each other, and the problem is inherently multi-objective with a trade-off.
\end{proof}

\textbf{Property 2}. \textit{If the driver discomfort only includes the traveling distance, excluding the observed queuing time, then minimizing driver discomfort is more inverse proportional to minimizing system inefficiency.}

\begin{proof}
Given the similar assumptions in the proof of Property 1, we set $\alpha_1 = 1$, $\alpha_2 = 0$, and assume that there exists a solution $x^{**}_{nm} = \arg \min \sum_{n=1}^N \sum_{m=1}^M D_{nm} \cdot x_{nm}$. When $t=0$, it is the same as the proof of Property 1. When $t>0$, EVs minimize their driver discomfort by choosing the nearest charging stations over all time, rather than consider the queuing time of stations. Therefore, the station $i$ in the traffic-congested areas is selected more than that $j$ in non-congested areas, $i,j \leq M$, i.e., $\sum_{n \in N_t} d_n \cdot x^{**}_{ni} > \sum_{n \in N_t} d_n \cdot x^{**}_{nj} = 0$. This also results in that the station $i$ is requested more frequently than the case that driver discomfort includes observed queuing time, i.e., $\sum_{n \in N_t} d_n \cdot x^{**}_{ni} > \sum_{n \in N_t} d_n \cdot x^*_{ni}$, where $x^*_{ni} = \arg \min [D_{ni} + \tau^{\text{Q}}_i(t)] \cdot x_{ni}$. As shown in Fig.~\ref{fig:appendix_prove}, the actual queuing times at station $i$ and $j$ are more imbalanced when the observed queuing time is not considered than when it is, leading to a higher $\mathcal{O}'_2$. Thus, the objective that minimizes discomfort is more inverse proportional to the objective that minimizes inefficiency, when the discomfort excludes the observed queuing time.
\end{proof}

\textbf{Property 3}. \textit{Real-time observation of charging station queuing times by EVs can alleviate the conflict between the two objectives, facilitating their simultaneous minimization.}

\begin{proof}
    Given the similar assumptions in the proof of Property 1, we allow all EVs coordinate once at the beginning by setting $T = 1$, and thus $x^{**}_{nm} = \arg \min \sum_{n=1}^N \alpha_1 \cdot \sum_{m=1}^M D_{nm} \cdot x_{nm}$. Based on the Property 2, the driver discomfort of EVs only includes the traveling distance, resulting in higher conflict between agent-centric and system-centric objectives.
\end{proof}

In conclusion, we formally prove that the objectives of driver discomfort and system inefficiency are contradictory to each other. This conflict is amplified if we exclude the observed queuing time from the driver discomfort. Without considering queuing delays, EVs greedily select the closest station, resulting in a skewed allocation and significantly increased variance in station load. This leads to a higher root mean square queuing time, which confirms the incompatibility of objectives and highlights the need for coordinating the selection behavior of EVs (i.e., autonomous charging behavior) in order to achieve Pareto optimality. 

\end{document}